\documentclass[aps,pre,reprint,twocolumn,longbibliography,floatfix,superscriptaddress]{revtex4}
\usepackage{graphicx}
\usepackage{amsmath,amssymb,amsfonts}
\usepackage{hyperref}
\begin{document}
\title[Random sequential adsorption of partially ordered discorectangles onto a continuous plane]{
Random sequential adsorption of partially ordered discorectangles onto a continuous plane}

\author{Nikolai I. Lebovka}
\email[Corresponding author: ]{lebovka@gmail.com}
\affiliation{Department of Physical Chemistry of Disperse Minerals, F. D. Ovcharenko Institute of Biocolloidal Chemistry, NAS of Ukraine, Kyiv 03142, Ukraine}
\affiliation{Department of Physics, Taras Shevchenko Kyiv National University, Kyiv 01033, Ukraine}
\author{Nikolai V. Vygornitskii}
\affiliation{Department of Physical Chemistry of Disperse Minerals, F. D. Ovcharenko Institute of Biocolloidal Chemistry, NAS of Ukraine, Kyiv, 03142 Ukraine}
\author{Yuri Yu. Tarasevich}
\email[Corresponding author: ]{tarasevich@asu.edu.ru}
\affiliation{Laboratory of Mathematical Modeling, Astrakhan State University, Astrakhan 414056, Russia}
\date{\today}

\begin{abstract}
Computer simulation was used to study the random sequential adsorption of identical discorectangles onto a continuous plane . The problem was analyzed for a wide range of discorectangle aspect ratios ($\varepsilon \in [1;100]$). We studied anisotropic deposition, i.e., the orientations of the deposited particles were uniformly distributed within some interval such that the particles were preferentially aligned along a given direction.
The kinetics of the changes in the packing density $\varphi$ found at different values of $S_0$ are discussed. Partial ordering of the discorectangles significantly affected the packing density at the jamming state, $\varphi_\text{j}$, and shifted the cusps in the $\varphi_\text{j}(\varepsilon)$ dependencies. The structure of the jammed state was analyzed using the adsorption of disks of different diameters into the porous space between the deposited discorectangles. The analysis of the connectivity between the discorectangles was performed assuming a core---shell structure of particles.
\end{abstract}

\maketitle

\section{Introduction\label{sec:intro}}

The behavior of systems of interacting elongated particles continues to attract great attention in both academic and applied fields. In such systems, complex collective behavior, spontaneous orientational ordering and self-assembly have been observed~\cite{Boerzsoenyi2013,Gan2020,Gan2020a}. In particular, systems of  elongated particles have shown a nematic orientational
ordering in both thermal equilibrium~\cite{Onsager1949,Bolhuis1997} and in athermal ($T=0$) systems (for example, uniform shear flow~\cite{Marschall2019}). The self-assembly of particles achieving their densest packing~\cite{Yu2006} and the crystallization transition (granular crystallization) from random to ordered packings  under mechanical vibration have been observed~\cite{Qian2019}. The particle shape may affect not only the packing characteristics of powders and granular materials, and of porous media (e.g., packing density and coordination numbers)~\cite{Zou1996,Guises2009,Kyrylyuk2011}, but also the processes of aggregation~\cite{Kwan2001}, gravity- and vibration-induced segregation~\cite{Abreu2003},  compression behavior~\cite{Azema2012}] and fluid flow through the porous packings~\cite{Chen1994}. A proper description of these processes is of fundamental importance for the preparation of advanced nanomaterials~\cite{Bokobza2019,Yang2019}, specifically those filled by nanotubes~\cite{Pampaloni2019} and nanoplatelets~\cite{Lebovka2019Springer}. The structure of packings filled by elongated particles can significantly affect the connectivity, electrical conductivity and permeation of such porous networks~\cite{Wang2020}.

In recent decades, the many important practical applications of thin films~\cite{Hirotani2019,Tiginyanu2019} have initiated great interest in studies of two-dimensional (2D) systems filled with elongated particles~\cite{Lebovka2020}. The rich phase behavior in these systems has been observed in its dependence on the confining dimension~\cite{Basurto2020}. Monte Carlo (MC) simulations of 2D systems of infinitely thin hard rods in thermal equilibrium have revealed a ``nematic'' phase at high densities~\cite{Frenkel1985}. However, such a ``nematic'' phase possessed algebraic order (quasi long-range order). For 2D fluids of discorectangles (rounded-cap rectangles)~\cite{Bates2000}, simulations have revealed that the ``nematic'' phase can only be observed for sufficiently long particles with aspect ratios ($\varepsilon$, length-to-width ratio) above 7. Shorter particles  do not exhibit a ``nematic'' phase, but undergo a melting transition.

Packing problems for non-equilibrium 2D systems of elongated particles have been intensively  studied using a random sequential adsorption (RSA) model~\cite{Evans1993,Adamczyk2017}. In this model, the particles are deposited randomly and sequentially onto a substrate, while overlapping with previously placed particles is strictly forbidden. For the RSA model, above some limiting coverage concentration $\varphi_\text{j}$ (called the jamming or saturation limit), there is no empty space for the deposition of a new particle and the adsorption process terminates.

The RSA simulations for disks gave a jamming coverage $\varphi_\text{j}=0.547 \pm 0.002$~\cite{Finegold1979Nat,Feder1980}. 2D saturated RSA packings of unoriented ellipses~\cite{Sherwood1990}, squares~\cite{Viot1990}, rectangles~\cite{Vigil1989,Vigil1990}, discorectangles~\cite{Haiduk2018},  polygons~\cite{Ciesla2014,Zhang2018}, sphere dimers, sphere polymers and other shapes~\cite{Ciesla2013,Ciesla2013a,Ciesla2015} have been investigated. For all the studied problems, cusp-like maximums of jamming coverage at some aspect ratios ($\varepsilon \approx 1.7-1.9$) were observed. For very elongated shapes ($\varepsilon \gg 1$), the value of $\varphi_\text{j}$ descended to zero according to the power law~\cite{Viot1992}:
\begin{equation}\label{eq:01}
	\varphi_\text{j} \propto \varepsilon^{-1/(1+\sqrt{2})}.
\end{equation}

Similar cusp-like maximums in the $\varphi_\text{j}(\varepsilon)$ dependencies have also been observed for saturated RSA packings of elongated particles in
one-dimensional (1D)~\cite{Chaikin2006,Baule2017,Ciesla2020,Lebovka2020a} and three-dimensional (3D)~\cite{Lebovka2020,Gan2020a} systems.

However, almost all previous studies of the saturated packing of elongated particles have been devoted to conventional RSA with unoriented particles. In some works, the RSA problems for perfectly oriented particles with respect to a selected direction, for example,
parallel squares~\cite{Brosilow1991} and ideally oriented superdisks~\cite{Gromenko2009} have been  studied. Recently, the thermal relaxation towards equilibrium of 2D oriented RSA packings have been investigated~\cite{Lebovka2019}. Here, the rods were infinitely thin ($\varepsilon = \infty$) and in the initial state, before relaxation, they were preferentially aligned with respect to a selected direction. The study revealed different relaxation behavior dependent on the preliminary ordering and the number density of the rods.

Experimental studies have reported  various effects of the alignment of elongated particles on the transport and optical properties of thin films~\cite{Ackermann2016,Wu2017}. Various alignment techniques have been proposed to organize elongated particles onto 2D substrates, based on external forces (magnetic~\cite{Shaver2009}, electrical~\cite{Mohammadimasoudi2016,Zande1999}), and shear flow~\cite{Wu2017}. This paper analyzes the RSA packing of identical elongated particles (discorectangles) on a 2D surface. We employ an off-lattice model, i.e., both the positions and orientations of the particles are continuous. The deposited particles were preferentially aligned along a selected direction. The aspect ratio of the particles varied within the range $\varepsilon \in [1;100]$. Special attention has been paid to the effects of the alignment of the particles on the packing characteristics.

The rest of the paper is constructed as follows. In Sec.~\ref{sec:methods}, the technical details of the simulations are described, all necessary quantities are defined, and some test results are presented. Section~\ref{sec:results} describes our principal findings. Section~\ref{sec:conclusion} summarizes the main results.

\section{Computational model\label{sec:methods}}

\begin{figure}[!htb]
  \centering
 \includegraphics[width=0.8\columnwidth]{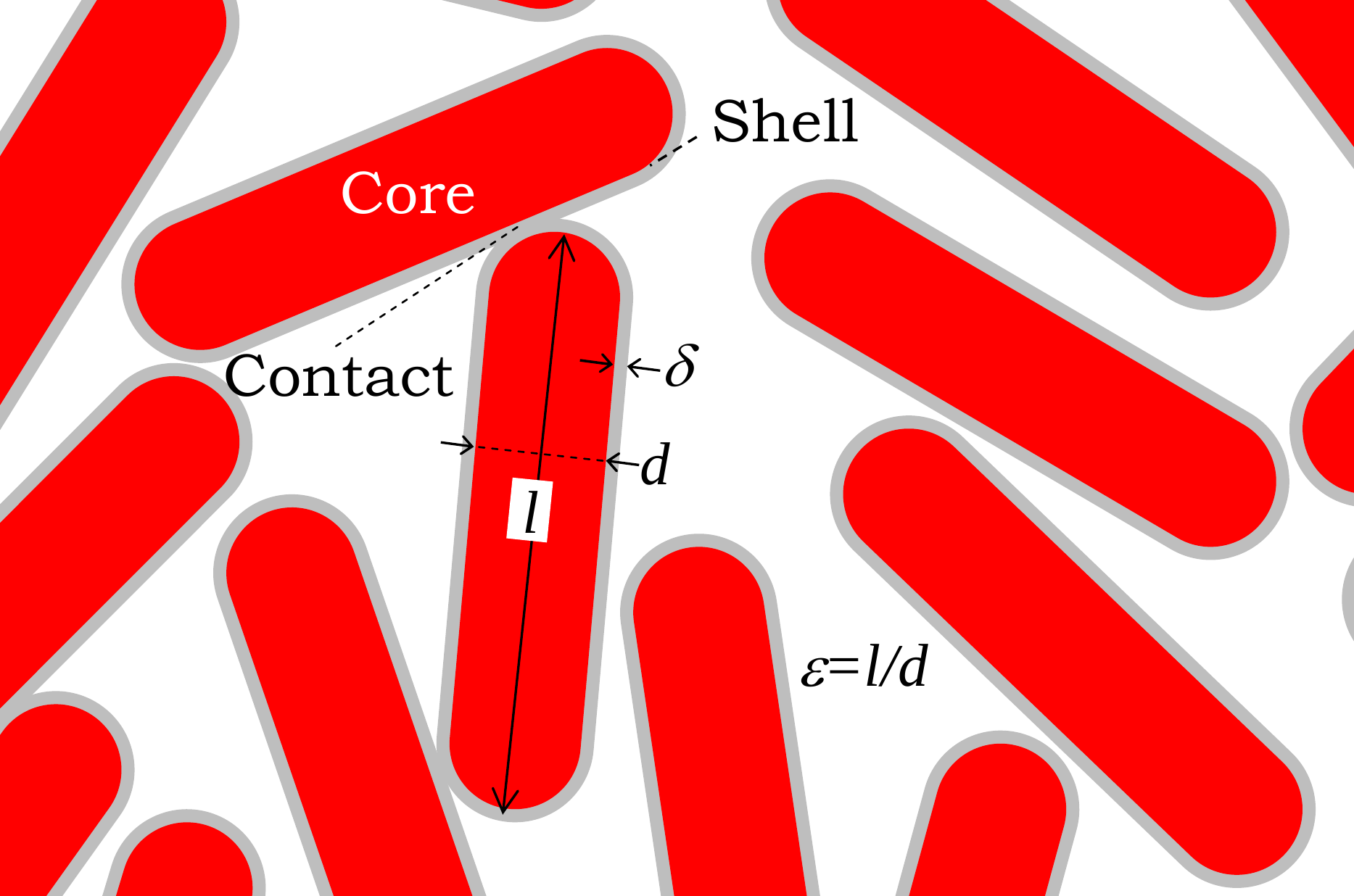}
  \caption{An illustration of the RSA model of the packing of discorectangles on a 2D substrate. Intersections of the particle cores are forbidden. For the connectivity analysis, each particle is assumed to be covered by a soft (penetrable) shell with thickness $\delta$.\label{fig:Connectf01}}
\end{figure}

A discorectangle is a rectangle with two hemidisks at its ends. The aspect ratio is defined as $\varepsilon=l/d$, where $l$ is the total length of the particle and $d$ is its width. The 2D packings were produced using an RSA model~\cite{Evans1993}. The particles were randomly and sequentially deposited onto a 2D surface until they reached the maximum (jamming) concentration $\varphi_\text{j}$. The overlapping of any new particle with previously deposited ones was forbidden (Fig.~\ref{fig:Connectf01}). For detecting overlapping during the deposition a fast algorithm to evaluate the shortest distance between particles was used~\cite{Vega1994,Pournin2005,Mahajan2018}. Periodic boundary conditions were applied to the substrate in both the $x$ and $y$ directions.

The preferential orientation of the particles  was characterized using the order parameter defined as
\begin{equation}\label{eq:S}
  S = \left\langle \cos 2\theta  \right\rangle,
\end{equation}
where $\langle\cdot\rangle$ denotes the average, $\theta$ is the angle between the long axis of the particle and the selected direction, and $x$ (the horizontal axis). For completely aligned and unoriented particles $S=1$, and $S=0$, respectively.

\begin{figure}[!htb]
  \centering
 \includegraphics[width=0.95\columnwidth]{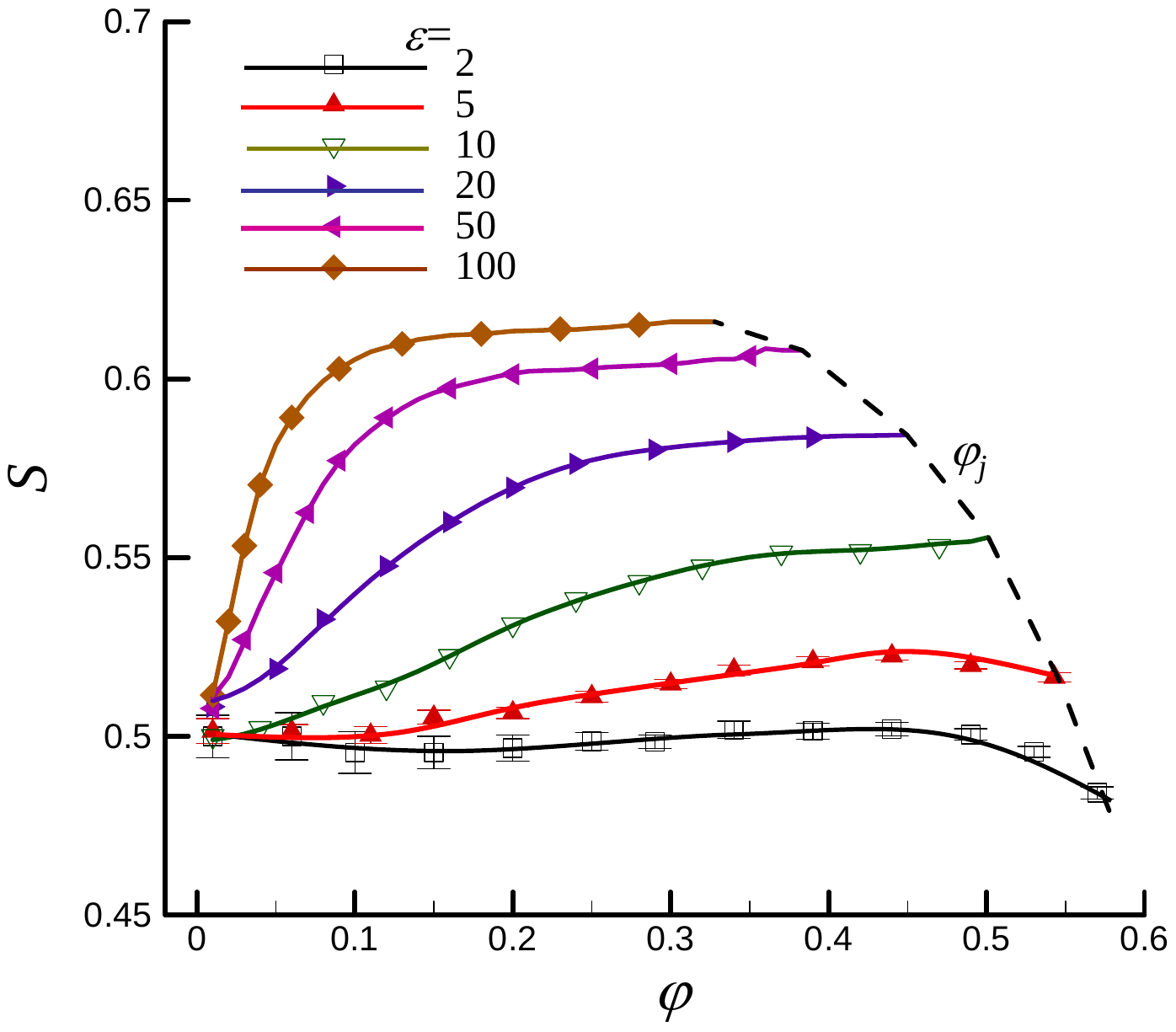}
  \caption{Actual order parameter, $S$, versus the packing density, $\varphi$, at different values of the aspect ratio $\varepsilon$. The preassigned order parameter is $S_0=0.5$. The values of $\varphi_\text{j}$  correspond to jamming states.
    \label{fig:02}}
\end{figure}

For producing partially ordered RSA deposits, a model of anisotropic random-orientation distribution has been used~\cite{Balberg1983}. For this model, the orientations of the deposited particles are selected to be uniformly distributed within some interval $\theta \in  [-\theta_\text{m};\theta_\text{m}]$, where $\theta_\text{m} \leqslant \pi/2$. In this case, the preassigned order parameter can be evaluated as~\cite{Lebovka2019}
\begin{equation}\label{eq:S0}
S_0 = \frac{\sin 2\theta_\text{m}}{2\theta_\text{m}}.
\end{equation}
The isotropic case $S_0=0$ corresponds to $\theta_\text{m}=\pi/2$. The smaller the value of $\theta_\text{m}$ is the higher the order parameter, $S_0$. During the deposition of particles according to the RSA protocol, some particle orientations may be rejected, therefore the actual order parameter in the deposit, $S$, may differ from the value of $S_0$. This situation resembles the RSA deposition of partially oriented elongated particles ($k$-mers) onto a square lattice~\cite{Lebovka2011}. The actual order parameter $S$ is conserved only for isotropic ($S_0=0$) and completely aligned ($S_0=1$) packing, while in the general case, the value of $S$ depends upon the packing density, $\varphi$. Figure~\ref{fig:02} presents examples of the actual order parameter, $S$, versus the packing density, $\varphi$, for the preassigned order parameter $S_0=0.5$ and different aspect ratios $\varepsilon$. Here, the values of $\varphi_\text{j}$  correspond to jamming states. When the aspect ratio is large, the difference between $S$ and $S_0$ may be fairly significant.

The dimensions of the system under consideration were $L$ along both the horizontal ($x$) and the vertical $y$ axes. In the present work, all calculations have been performed using $L=32 l$. For each given value of $\varepsilon$ or $S_0$, the computer  experiments were repeated up to $100$ times.  The error bars in the figures correspond to the standard deviation of the mean. When not shown explicitly, they are of the order of the marker size. All simulations were performed for the aspect ratios $\varepsilon \in [1;100]$.

In the resulting deposits, the particles cannot touch each other, hence, they are not in direct contact. However, a core---shell model of the particles can be used to evaluate the connectedness of the particles. To perform this analysis, each particle was covered by an outer shell with a thickness $\delta$ (Fig.~\ref{fig:Connectf01}). Any two particles are assumed to be connected when the minimal distance between their hard cores does not exceed the value of $2\delta$.

The minimum (critical) value of the outer shell thickness $\delta$ required for the formation of spanning clusters along the $x$ or $y$ direction, was evaluated. The calculations were performed using the Hoshen---Kopelman algorithm~\cite{Hoshen1976}. The analysis was carried out using a list of near-neighbor particles~\cite{Marck1997}.

\begin{figure}[!htb]
  \centering
 \includegraphics[width=0.95\columnwidth]{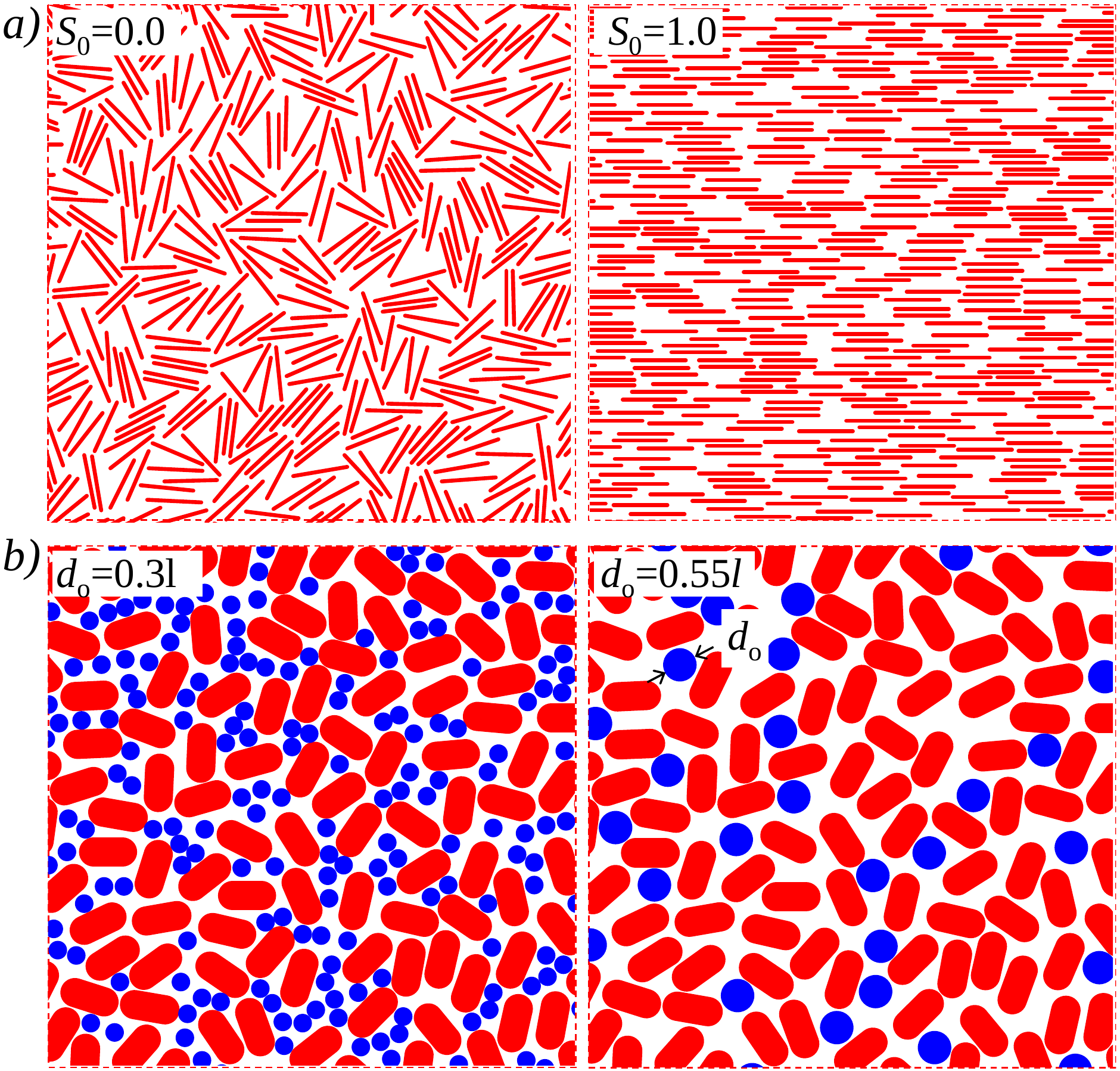}
  \caption{Examples of the jamming patterns for aspect ratio $\varepsilon=10$  at $S_0=0$ (random orientation) and $S_0=1$ (complete alignment along the horizontal direction) (a), and for aspect ratio $\varepsilon=2$ at $S_0=0$ (b). For the latter case, the void space between the discorectangles is filled with reference disks with diameters $d_\circ=0.3l$ and $d_\circ=0.55l$. Fragments with size of $9l\times 9l$ are shown. \label{fig:Patternsf03}
  }
\end{figure}

Figure~\ref{fig:Patternsf03}a presents examples of the jamming  patterns for a fixed aspect ratio of $\varepsilon=10$  at $S_0=0$ (random orientation) and $S_0=1$ (complete alignment along the horizontal direction $x$). An analysis of the holes between discorectangles was undertaken using reference disks of different diameters, $d_\circ$. In these tests, after the formation of the jamming deposits, the ``accessible void'' was supplementarily filled with the reference disks using the RSA model up to the jamming limit. Figure~\ref{fig:Patternsf03}b presents examples of the jamming patterns  for the aspect ratio  $\varepsilon=2$  and random orientation of the discorectangles ($S_0=0$) for $d_\circ=0.3l$ and $d_\circ=0.55l$.


\section{Results and Discussion\label{sec:results}}
\begin{figure}[!htb]
  \centering
 \includegraphics[width=0.95\columnwidth]{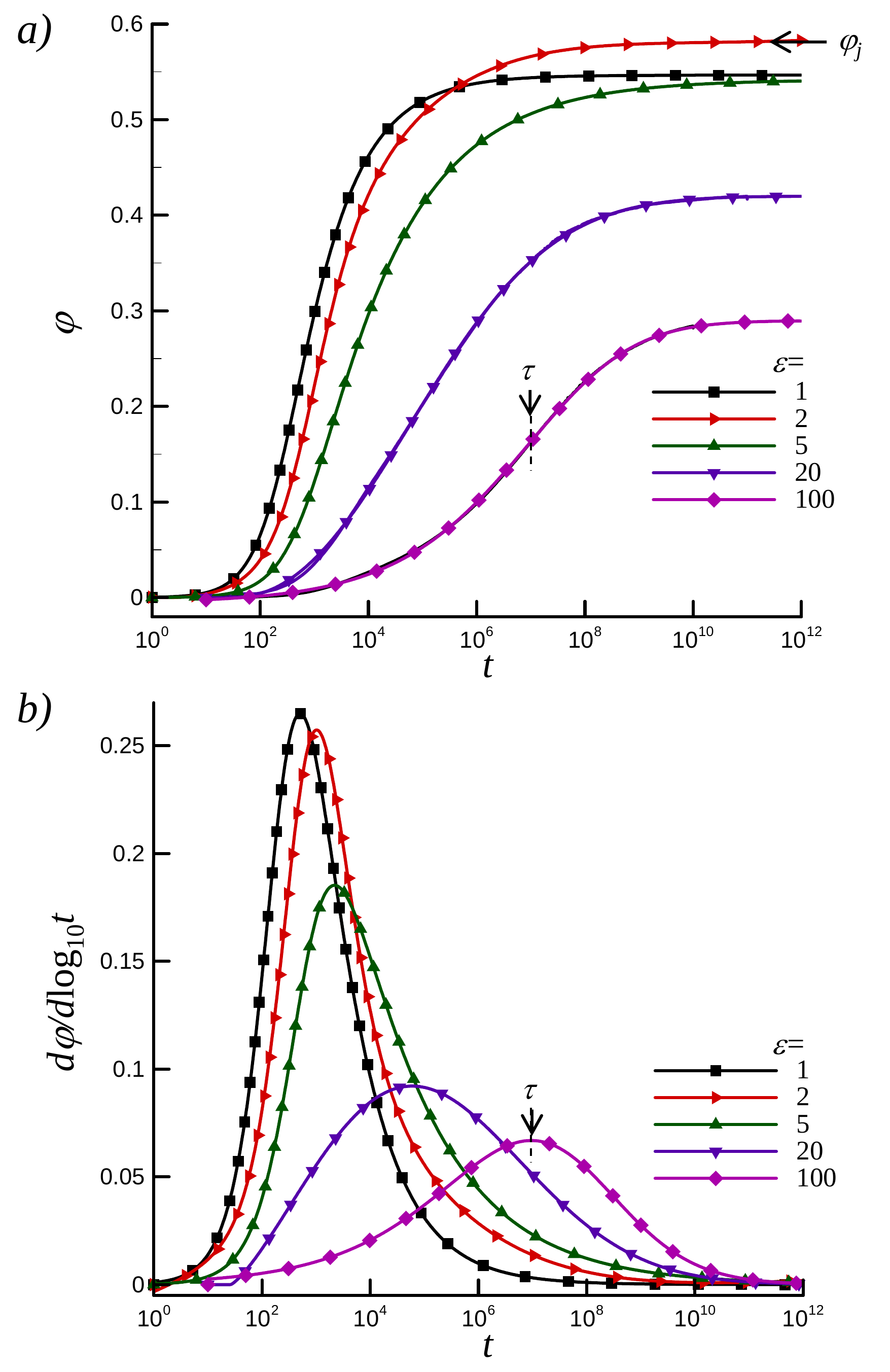}
  \caption{Packing density,  $\varphi$, (a) and time derivative $\mathrm{d} \varphi/\mathrm{d}\log_{10} t$ (b) versus the deposition time, $t$, for the disordered RSA packing ($S_0=0$) of discorectangles with aspect ratio $\varepsilon$. Here, $\varphi_\text{j}$ is the jamming concentration and $\tau$ is the characteristic deposition time. \label{fig:f04}}
\end{figure}

Figure~\ref{fig:f04}a presents the packing density,  $\varphi$, versus the deposition time, $t$, for the disordered RSA packing ($S_0=0$) of discorectangles with different aspect ratios $\varepsilon$. The value of $\varphi$ gradually increased with increasing $t$ as the system approached the jamming value $\varphi_\text{j}$ at $t \to \infty$. Similar dependencies $\varphi (t)$ were also observed for other values of $S_0$. The inflections in the time derivatives $\mathrm{d} \varphi/\mathrm{d}\log_{10} t$ were used to estimate the characteristic deposition times, $\tau$ (Fig.~\ref{fig:f04}b)~\cite{Hart2016PRE}.

\begin{figure}[!htb]
  \centering
 \includegraphics[width=0.95\columnwidth]{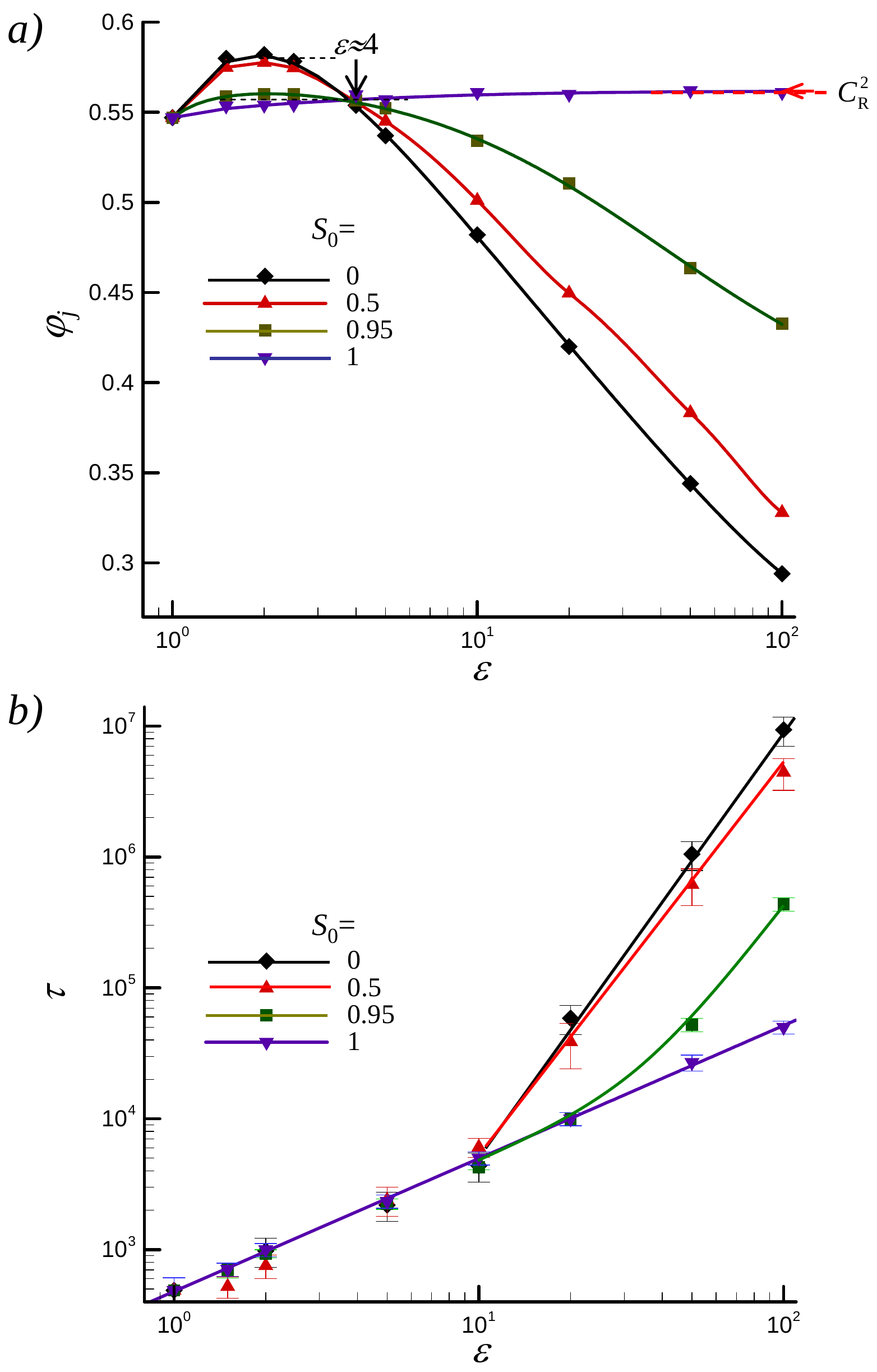}
  \caption{Packing density at the jamming state, $\varphi_\text{j}$, (a) and characteristic deposition time,
  $\tau$, (b) versus the discorectangle aspect ratio, $\varepsilon$, at different values of the preassigned order parameter, $S_0$.
  \label{fig:f05}}
\end{figure}

Figure~\ref{fig:f05}a shows the packing density at the jamming state, $\varphi_\text{j}$, versus the discorectangle aspect ratio, $\varepsilon$, at different values of the preassigned order parameter, $S_0$. For partially disordered systems (at $S_0<1$) noticeable cusps (maximums) in the $\varphi_\text{j}(\varepsilon)$ dependencies could be observed. For example, for completely disordered RSA packing ($S_0 =0$) a well-defined maximum $\varphi_\text{j}= 0.583 \pm 0.004$ (at $\varepsilon  \approx 1.46$) was observed and this result is in good correspondence with a previously reported value for completely disordered discorectangles,
$\varphi_\text{j} = 0.582896 \pm 0.000019 $~\cite{Haiduk2018}. The initial density increase can be explained by relaxing the parameter constraint (appearance of orientational degrees of freedom) in the RSA packing of the elongated particles, while  the density decrease at larger values of $\varepsilon$ may reflect excluded volume effects~\cite{Chaikin2006}. An increase of partial ordering noticeably influenced the character of the $\varphi_\text{j}(\varepsilon)$ dependencies and the cusps became less significant. A remarkable feature was the presence a stable point at $\varepsilon\simeq 4$ with nearly the same values of $\varphi_\text{j}=0.557\pm 0.002$ for all values of $S_0 \in [0;1]$. In the limit of completely aligned discorectangles ($S_0 \to 1$) the cusp disappeared. For this case, the packing density gradually increased with $\varepsilon$ and approached the value $\varphi_\text{j} \approx C^2_\text{R} = 0.5589\dots $
(Fig.~\ref{fig:f05}a), where  $C_\text{R}$ is the well-known R\'{e}nyi's parking constant for a 1D problem~\cite{Renyi1963}. This supports Pal\'{a}sti's conjecture, regarding the relationships of the jamming limits for 1D  and 2D problems~\cite{Palasti1960}.

Figure~\ref{fig:f05}b shows the characteristic deposition time, $\tau$, versus the discorectangle aspect ratio, $\varepsilon$, at different values of the preassigned order parameter, $S_0$. At relatively small aspect ratios ($\varepsilon \leqslant 10$) an approximately linear increase of $\tau$ with $\varepsilon$ for all values of $S_0$ was observed:
\begin{equation}\label{eq:tau}
  \tau=\tau_1+a(\varepsilon-1),
\end{equation}
where $\tau_1= 490\pm 12$ corresponds to the characteristic deposition time for disks ($\varepsilon=1$) and
$a= 508 \pm	6$.

For completely aligned systems ($S_0=1$), this linear $\tau(\varepsilon)$ dependence was also observed for $\varepsilon > 10$. For partially oriented systems ($S_0<1$), significant deviations were observed for $\varepsilon > 10$, particularly for $S_0=0$. This reflected the excluded volume effects on stagnation of the RSA deposition process for disordered systems.

\begin{figure}[!htb]
  \centering
 \includegraphics[width=0.95\columnwidth]{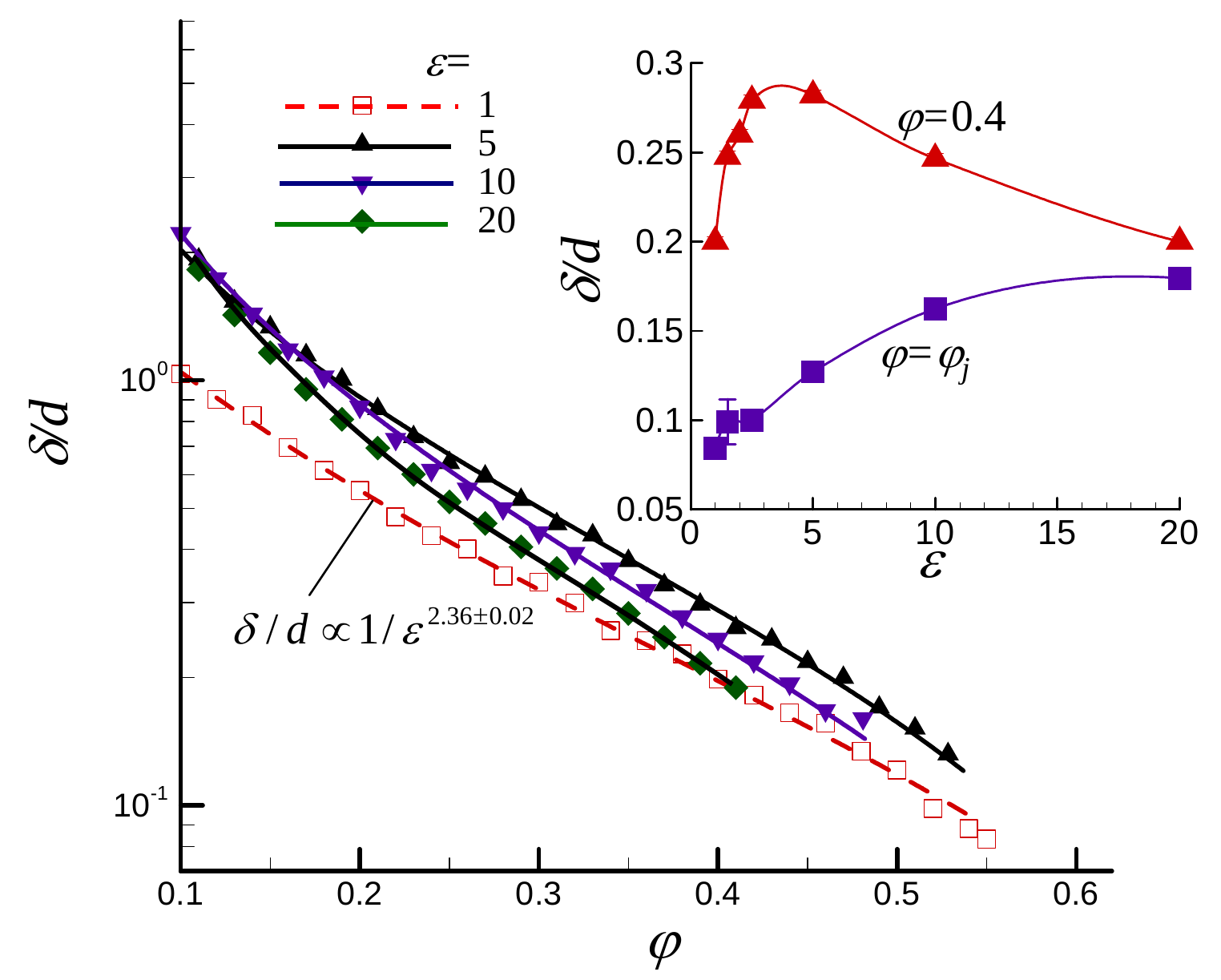}
  \caption{Minimal reduced thickness of shell, $\delta/d$, required for infinite connectivity between particles (formation of a spanning path through the system), versus the packing density, $\varphi$, at different aspect ratios $\varepsilon$. The preassigned order parameter is $S_0=0.0$. The inset shows the $\delta/d$ versus $\varepsilon$ dependencies at $\varphi =0.4$ and $\varphi=\varphi_\text{j}$ (jamming state).
  \label{fig:f06}}
\end{figure}

Figure~\ref{fig:f06} demonstrates examples of the minimal reduced thickness of the shell, $\delta/d$, required for a spanning path through the system, versus the packing density, $\varphi$, at a fixed preassigned order parameter $S_0=0.0$ and different aspect ratios $\varepsilon$. The values of $\delta/d$ decreased with $\varphi$, particularly, the power relation $\delta/d \propto 1/\varphi^{\alpha}$ ($\alpha = 2.36\pm 0.02$) could be observed at $\varepsilon=1$. However, the connectivity behavior at different values of $\varepsilon$ was rather complex (see inset in  Fig.~\ref{fig:f06}). For example, at a fixed value of $\varphi=0.4$, the connectivity analysis revealed a cusp (maximum) in the $\delta(\varepsilon)/d$ dependence. This maximum is similar to that observed in the $\varphi_\text{j}/(\varepsilon)$ dependence and it evidently reflects the interplay of the above-mentioned geometrical effects (orientation freedom and excluded volume). For the maximum packing (jamming state) the value of $\delta/d$ increased with $\varepsilon$, i.e., the connectivity did not display any cusp behavior (see inset in Fig.~\ref{fig:f06}).

\begin{figure}[!htb]
  \centering
 \includegraphics[width=0.95\columnwidth]{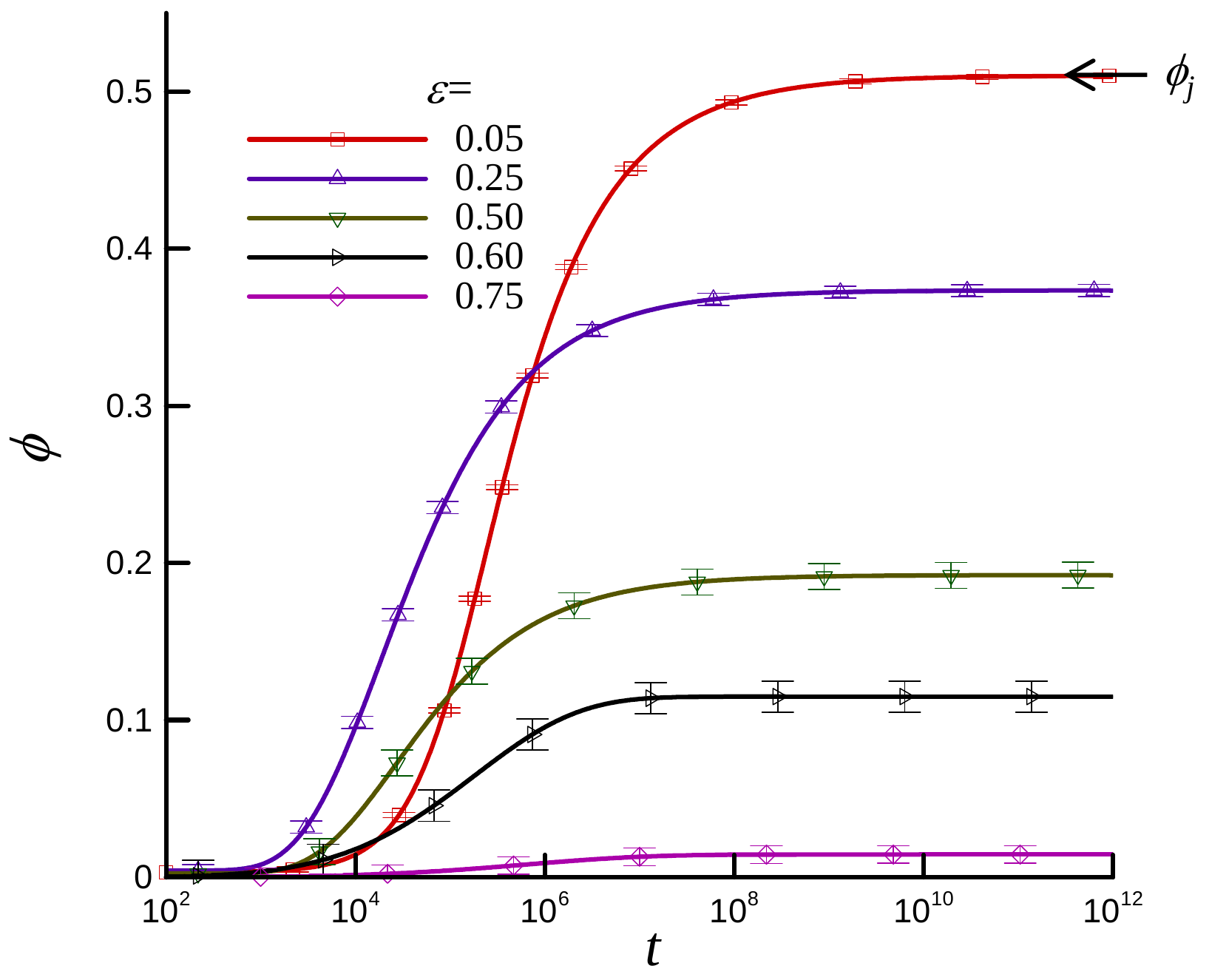}
  \caption{Packing density of reference disks, $\phi$, versus the deposition time, $t$. Here,  discorectangles with an aspect ratio of $\varepsilon=2$ and random orientations ($S_0=0$) were preliminarily deposited to the jamming state ($\varphi_ \text{j} \approx 0.582$), and then RSA packing of reference disks with different diameters $d_\circ$ into the confined void spaces between the discorectangles was applied. The value $\phi_\text{j}$ corresponds to the jamming concentration of the reference disks.\label{fig:f07}}
\end{figure}

Figure~\ref{fig:f07} demonstrates examples of the packing density of reference disks, $\phi$, versus the deposition time, $t$. In these simulations, discorectangles with an aspect ratio of $\varepsilon=2$ and random orientations ($S_0=0$) were preliminarily deposited to the jamming state ($\varphi_ \text{j} \approx 0.582$). Then RSA packing of reference disks with different diameters $d_\circ$ into the confined void spaces between the discorectangles was applied. The value $\phi_\text{j}$ corresponds to the jamming concentration of the reference disks, and with small disks ($d_0/l\ll 1$), it was close to the value of $\phi_\text{j}\approx 0.547$ for the jamming concentration of disks seen for an non-confined RSA problem on a plane~\cite{Finegold1979Nat,Feder1980}.

\begin{figure}[!htb]
  \centering
 \includegraphics[width=0.95\columnwidth]{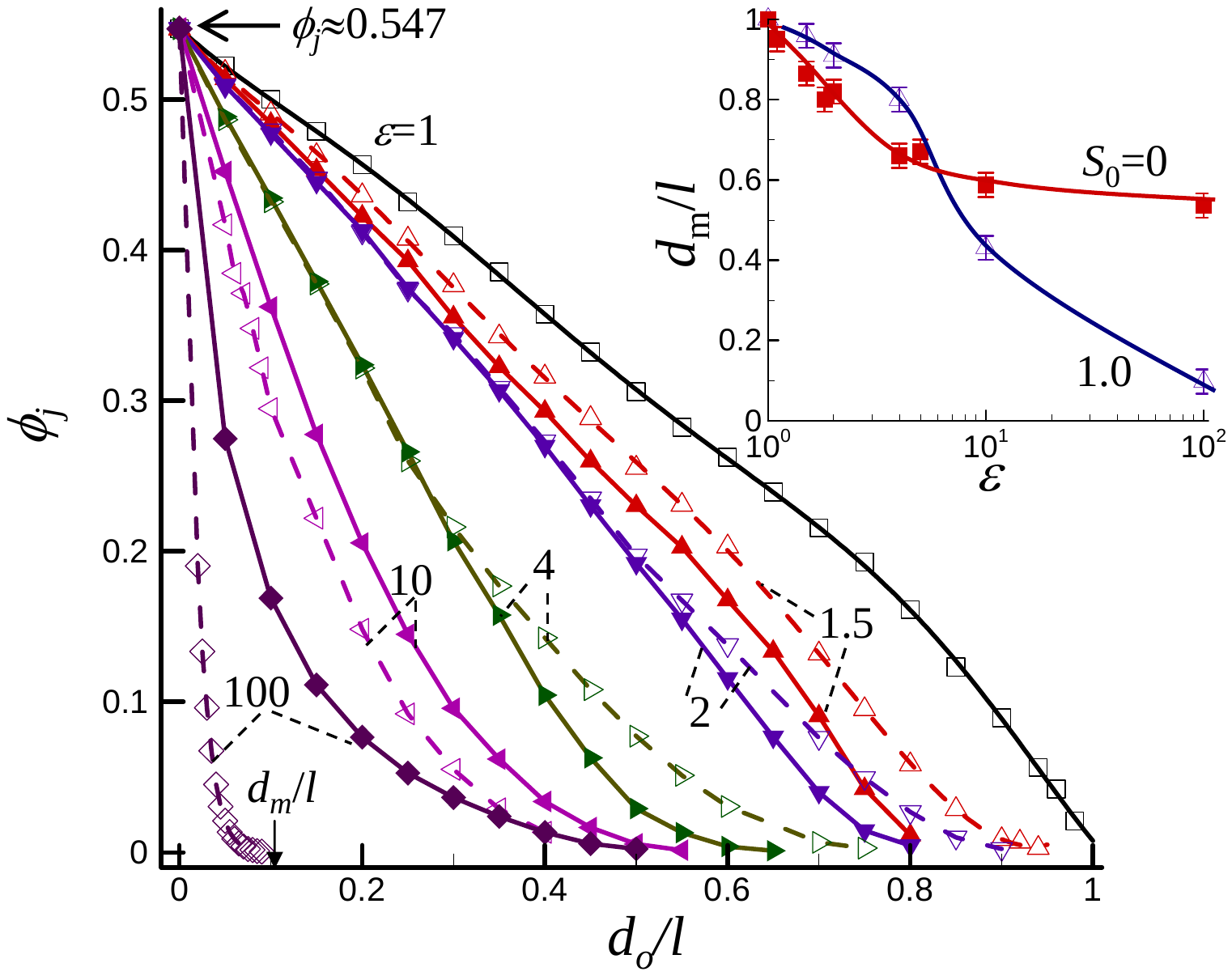}
  \caption{Jamming packing density of reference disks in the void space between discorectangles, $\phi_\text{j}$, versus reduced diameter of the disks $d_\circ/l$. The data are presented for completely disordered ($S_0=0$, filled symbols, solid lines) and completely aligned ($S_0=1$, open symbols, dashed lines) discorectangles with different aspect ratios of $\varepsilon$. The value of $d_\text{m}/l$ corresponds to the maximum diameter of the reference disks. Inset shows the $d_\text{m}/l$ versus $\varepsilon$ dependencies for different values of the preassigned order parameters, $S_0$.\label{fig:f08}}
\end{figure}
Figure~\ref{fig:f08} compares the $\phi_\text{j}$ versus $d_\circ/l$ dependencies for completely disordered ($S_0=0$, filled symbols, solid lines) and completely aligned ($S_0=1$, open symbols, dashed lines) discorectangles at different aspect ratios, $\varepsilon$. The values of $\phi_\text{j}$ gradually decreased with $d_\circ/l$ and where RSA packing was possible for reference disks with sizes not exceeding some maximum size $d_\text{m}/l$. The character of the $\phi_\text{j}(d_\circ/l)$ dependencies were noticeably different for different values of  $\varepsilon$. They were approximately linear at some intermediate values of the aspect ratio $\varepsilon \approx 2$, demonstrated convexity when $\varepsilon < 2$, and concavity when $\varepsilon > 2$. Moreover, the value of $\varepsilon$ differently affected the maximum diameter $d_\text{m}/l$ in the dependence on $S_0$ (see inset in Fig.~\ref{fig:f07}). At small values of $\varepsilon$ ($\varepsilon\lessapprox 4$) the value of $d_\text{m}/l$ for a completely aligned system ($S_0=1$) exceeded that for a completely disordered system ($S_0=0$), but the situation was inverse at large values of $\varepsilon$ ($\varepsilon> 4$). For large aspect ratios ($\varepsilon\gg 1$) a different limiting behavior was also observed, $d_\text{m}/l \rightarrow 0.5$ for $S_0=0$ and  $d_\text{m}/l \rightarrow 0$ for $S_0=1$. For relatively large values of $\varepsilon$ the relatively large holes (``empty'' spaces) between the discorectangles  can be distinguished accounting for the side-to-side, side-to-cap and cap-to-cap positions of the near-neighbor discorectangles (Fig.~\ref{fig:Patternsf03}). This distinction is governed by the preassigned order parameter, $S_0$, and a decreased value of this results in an increase in the  size of the holes (value of $d_\text{m}/l$).  Note, that  side-to-side contacts were expected to play a stabilizing role in the packing of elongated particles~\cite{Azema2010}.

\section{Conclusion\label{sec:conclusion}}

Simulations for a continuous 2D model of RSA packing of discorectangles have been performed. The initial state was produced at different values of the preassigned order parameters $0 \leqslant S_0\leqslant 1$. The effects of ordering on the packings were noticeably dependent on the discorectangle aspect ratio, $\varepsilon$.  For partially disordered systems ($S_0<1$) and at relatively small values of $\varepsilon$ ($\varepsilon <4$), the dependencies of the jamming density $\varphi_\text{j}(\varepsilon)$ showed that the cusps and the values of $\varphi_\text{j}$ decreased with $\varepsilon$ for  more elongated particles ($\varepsilon > 4$). However, for completely aligned discorectangles ($S_0= 1$) the cusps disappeared, and the value of $\varphi_\text{j}$ gradually increased with $\varepsilon$ and approached to the value $\varphi_\text{j} \approx C^2_\text{R}$, where  $C_\text{R}$ is the R\'{e}nyi's parking constant for a 1D problem~\cite{Renyi1963}. Therefore, Pal\'{a}sti's conjecture, that jamming coverages in 1D and 2D are connected as ($\varphi_\text{j,2D}= \varphi_\text{j,1D}^2$), is fulfilled~\cite{Palasti1960}. At $\varepsilon=4$,  we observed almost the same value of $\varphi_\text{j}=0.557\pm 0.002)$ for all values of $S_0 \in [0;1]$. The complex effects of aspect ratio and orientation ordering on the connectivity of discorectangles with core---shell structures  and on the distribution of local voids between discorectangles were also revealed.

\acknowledgments
We acknowledge funding from the National Academy of Sciences of Ukraine, Projects No.~0117U004046 and  Project within the program ``Perspective fundamental research and innovative development of nanomaterials and nanotechnologies for the needs of industry, health care and agriculture'' (2020--2024) (N.I.L., N.V.V.), and the Russian Foundation for Basic Research, Project No.~18-07-00343 (Yu.Yu.T.).

\bibliography{Packing2dshort}

\begin{thebibliography}{64}%
\makeatletter
\providecommand \@ifxundefined [1]{%
 \@ifx{#1\undefined}
}%
\providecommand \@ifnum [1]{%
 \ifnum #1\expandafter \@firstoftwo
 \else \expandafter \@secondoftwo
 \fi
}%
\providecommand \@ifx [1]{%
 \ifx #1\expandafter \@firstoftwo
 \else \expandafter \@secondoftwo
 \fi
}%
\providecommand \natexlab [1]{#1}%
\providecommand \enquote  [1]{``#1''}%
\providecommand \bibnamefont  [1]{#1}%
\providecommand \bibfnamefont [1]{#1}%
\providecommand \citenamefont [1]{#1}%
\providecommand \href@noop [0]{\@secondoftwo}%
\providecommand \href [0]{\begingroup \@sanitize@url \@href}%
\providecommand \@href[1]{\@@startlink{#1}\@@href}%
\providecommand \@@href[1]{\endgroup#1\@@endlink}%
\providecommand \@sanitize@url [0]{\catcode `\\12\catcode `\$12\catcode
  `\&12\catcode `\#12\catcode `\^12\catcode `\_12\catcode `\%12\relax}%
\providecommand \@@startlink[1]{}%
\providecommand \@@endlink[0]{}%
\providecommand \url  [0]{\begingroup\@sanitize@url \@url }%
\providecommand \@url [1]{\endgroup\@href {#1}{\urlprefix }}%
\providecommand \urlprefix  [0]{URL }%
\providecommand \Eprint [0]{\href }%
\providecommand \doibase [0]{https://doi.org/}%
\providecommand \selectlanguage [0]{\@gobble}%
\providecommand \bibinfo  [0]{\@secondoftwo}%
\providecommand \bibfield  [0]{\@secondoftwo}%
\providecommand \translation [1]{[#1]}%
\providecommand \BibitemOpen [0]{}%
\providecommand \bibitemStop [0]{}%
\providecommand \bibitemNoStop [0]{.\EOS\space}%
\providecommand \EOS [0]{\spacefactor3000\relax}%
\providecommand \BibitemShut  [1]{\csname bibitem#1\endcsname}%
\let\auto@bib@innerbib\@empty
\bibitem [{\citenamefont {B{\"o}rzs{\"o}nyi}\ and\ \citenamefont
  {Stannarius}(2013)}]{Boerzsoenyi2013}%
  \BibitemOpen
  \bibfield  {author} {\bibinfo {author} {\bibfnamefont {T.}~\bibnamefont
  {B{\"o}rzs{\"o}nyi}}\ and\ \bibinfo {author} {\bibfnamefont {R.}~\bibnamefont
  {Stannarius}},\ }\bibfield  {title} {\bibinfo {title} {Granular materials
  composed of shape-anisotropic grains},\ }\href
  {https://doi.org/10.1039/C3SM50298H} {\bibfield  {journal} {\bibinfo
  {journal} {Soft Matter}\ }\textbf {\bibinfo {volume} {9}},\ \bibinfo {pages}
  {7401} (\bibinfo {year} {2013})}\BibitemShut {NoStop}%
\bibitem [{\citenamefont {Gan}\ and\ \citenamefont
  {Yu}(2020{\natexlab{a}})}]{Gan2020}%
  \BibitemOpen
  \bibfield  {author} {\bibinfo {author} {\bibfnamefont {J.}~\bibnamefont
  {Gan}}\ and\ \bibinfo {author} {\bibfnamefont {A.}~\bibnamefont {Yu}},\
  }\bibfield  {title} {\bibinfo {title} {{DEM} study on the packing density and
  randomness for packing of ellipsoids},\ }\href
  {https://doi.org/10.1016/j.powtec.2019.07.012} {\bibfield  {journal}
  {\bibinfo  {journal} {Powder Technol.}\ }\textbf {\bibinfo {volume} {361}},\
  \bibinfo {pages} {424} (\bibinfo {year} {2020}{\natexlab{a}})}\BibitemShut
  {NoStop}%
\bibitem [{\citenamefont {Gan}\ and\ \citenamefont
  {Yu}(2020{\natexlab{b}})}]{Gan2020a}%
  \BibitemOpen
  \bibfield  {author} {\bibinfo {author} {\bibfnamefont {J.}~\bibnamefont
  {Gan}}\ and\ \bibinfo {author} {\bibfnamefont {A.}~\bibnamefont {Yu}},\
  }\bibfield  {title} {\bibinfo {title} {{DEM} simulation of the packing of
  cylindrical particles},\ }\href {https://doi.org/10.1007/s10035-019-0993-4}
  {\bibfield  {journal} {\bibinfo  {journal} {Granul. Matter}\ }\textbf
  {\bibinfo {volume} {22}},\ \bibinfo {pages} {1} (\bibinfo {year}
  {2020}{\natexlab{b}})}\BibitemShut {NoStop}%
\bibitem [{\citenamefont {Onsager}(1949)}]{Onsager1949}%
  \BibitemOpen
  \bibfield  {author} {\bibinfo {author} {\bibfnamefont {L.}~\bibnamefont
  {Onsager}},\ }\bibfield  {title} {\bibinfo {title} {The effects of shape on
  the interaction of colloidal particles},\ }\href
  {https://doi.org/10.1111/j.1749-6632.1949.tb27296.x} {\bibfield  {journal}
  {\bibinfo  {journal} {Ann. N.Y. Acad. Sci.}\ }\textbf {\bibinfo {volume}
  {51}},\ \bibinfo {pages} {627} (\bibinfo {year} {1949})}\BibitemShut
  {NoStop}%
\bibitem [{\citenamefont {Bolhuis}\ and\ \citenamefont
  {Frenkel}(1997)}]{Bolhuis1997}%
  \BibitemOpen
  \bibfield  {author} {\bibinfo {author} {\bibfnamefont {P.}~\bibnamefont
  {Bolhuis}}\ and\ \bibinfo {author} {\bibfnamefont {D.}~\bibnamefont
  {Frenkel}},\ }\bibfield  {title} {\bibinfo {title} {Tracing the phase
  boundaries of hard spherocylinders},\ }\href
  {https://doi.org/10.1063/1.473404} {\bibfield  {journal} {\bibinfo  {journal}
  {J. Chem. Phys.}\ }\textbf {\bibinfo {volume} {106}},\ \bibinfo {pages} {666}
  (\bibinfo {year} {1997})}\BibitemShut {NoStop}%
\bibitem [{\citenamefont {Marschall}\ \emph {et~al.}(2019)\citenamefont
  {Marschall}, \citenamefont {Keta}, \citenamefont {Olsson},\ and\
  \citenamefont {Teitel}}]{Marschall2019}%
  \BibitemOpen
  \bibfield  {author} {\bibinfo {author} {\bibfnamefont {T.}~\bibnamefont
  {Marschall}}, \bibinfo {author} {\bibfnamefont {Y.-E.}\ \bibnamefont {Keta}},
  \bibinfo {author} {\bibfnamefont {P.}~\bibnamefont {Olsson}},\ and\ \bibinfo
  {author} {\bibfnamefont {S.}~\bibnamefont {Teitel}},\ }\bibfield  {title}
  {\bibinfo {title} {Orientational ordering in athermally sheared, aspherical,
  frictionless particles},\ }\href
  {https://doi.org/10.1103/PhysRevLett.122.188002} {\bibfield  {journal}
  {\bibinfo  {journal} {Phys. Rev. Lett.}\ }\textbf {\bibinfo {volume} {122}},\
  \bibinfo {pages} {188002} (\bibinfo {year} {2019})}\BibitemShut {NoStop}%
\bibitem [{\citenamefont {Yu}\ \emph {et~al.}(2006)\citenamefont {Yu},
  \citenamefont {An}, \citenamefont {Zou}, \citenamefont {Yang},\ and\
  \citenamefont {Kendall}}]{Yu2006}%
  \BibitemOpen
  \bibfield  {author} {\bibinfo {author} {\bibfnamefont {A.~B.}\ \bibnamefont
  {Yu}}, \bibinfo {author} {\bibfnamefont {X.~Z.}\ \bibnamefont {An}}, \bibinfo
  {author} {\bibfnamefont {R.~P.}\ \bibnamefont {Zou}}, \bibinfo {author}
  {\bibfnamefont {R.~Y.}\ \bibnamefont {Yang}},\ and\ \bibinfo {author}
  {\bibfnamefont {K.}~\bibnamefont {Kendall}},\ }\bibfield  {title} {\bibinfo
  {title} {Self-assembly of particles for densest packing by mechanical
  vibration},\ }\href {https://doi.org/10.1103/PhysRevLett.97.265501}
  {\bibfield  {journal} {\bibinfo  {journal} {Phys. Rev. Lett.}\ }\textbf
  {\bibinfo {volume} {97}},\ \bibinfo {pages} {265501} (\bibinfo {year}
  {2006})}\BibitemShut {NoStop}%
\bibitem [{\citenamefont {Qian}\ \emph {et~al.}(2019)\citenamefont {Qian},
  \citenamefont {An}, \citenamefont {Zhao}, \citenamefont {Dong}, \citenamefont
  {Wu}, \citenamefont {Fu}, \citenamefont {Zhang},\ and\ \citenamefont
  {Yang}}]{Qian2019}%
  \BibitemOpen
  \bibfield  {author} {\bibinfo {author} {\bibfnamefont {Q.}~\bibnamefont
  {Qian}}, \bibinfo {author} {\bibfnamefont {X.}~\bibnamefont {An}}, \bibinfo
  {author} {\bibfnamefont {H.}~\bibnamefont {Zhao}}, \bibinfo {author}
  {\bibfnamefont {K.}~\bibnamefont {Dong}}, \bibinfo {author} {\bibfnamefont
  {Y.}~\bibnamefont {Wu}}, \bibinfo {author} {\bibfnamefont {H.}~\bibnamefont
  {Fu}}, \bibinfo {author} {\bibfnamefont {H.}~\bibnamefont {Zhang}},\ and\
  \bibinfo {author} {\bibfnamefont {X.}~\bibnamefont {Yang}},\ }\bibfield
  {title} {\bibinfo {title} {Particle scale study on the crystallization of
  mono-sized cylindrical particles subject to vibration},\ }\href
  {https://doi.org/10.1016/j.powtec.2019.05.002} {\bibfield  {journal}
  {\bibinfo  {journal} {Powder Technol.}\ }\textbf {\bibinfo {volume} {352}},\
  \bibinfo {pages} {470} (\bibinfo {year} {2019})}\BibitemShut {NoStop}%
\bibitem [{\citenamefont {Zou}\ and\ \citenamefont {Yu}(1996)}]{Zou1996}%
  \BibitemOpen
  \bibfield  {author} {\bibinfo {author} {\bibfnamefont {R.~P.}\ \bibnamefont
  {Zou}}\ and\ \bibinfo {author} {\bibfnamefont {A.~B.}\ \bibnamefont {Yu}},\
  }\bibfield  {title} {\bibinfo {title} {Evaluation of the packing
  characteristics of mono-sized non-spherical particles},\ }\href
  {https://doi.org/10.1016/0032-5910(96)03106-3} {\bibfield  {journal}
  {\bibinfo  {journal} {Powder Technol.}\ }\textbf {\bibinfo {volume} {88}},\
  \bibinfo {pages} {71} (\bibinfo {year} {1996})}\BibitemShut {NoStop}%
\bibitem [{\citenamefont {Guises}\ \emph {et~al.}(2009)\citenamefont {Guises},
  \citenamefont {Xiang}, \citenamefont {Latham},\ and\ \citenamefont
  {Munjiza}}]{Guises2009}%
  \BibitemOpen
  \bibfield  {author} {\bibinfo {author} {\bibfnamefont {R.}~\bibnamefont
  {Guises}}, \bibinfo {author} {\bibfnamefont {J.}~\bibnamefont {Xiang}},
  \bibinfo {author} {\bibfnamefont {J.-P.}\ \bibnamefont {Latham}},\ and\
  \bibinfo {author} {\bibfnamefont {A.}~\bibnamefont {Munjiza}},\ }\bibfield
  {title} {\bibinfo {title} {Granular packing: numerical simulation and the
  characterisation of the effect of particle shape},\ }\href
  {https://doi.org/10.1007/s10035-009-0148-0} {\bibfield  {journal} {\bibinfo
  {journal} {Granul. Matter}\ }\textbf {\bibinfo {volume} {11}},\ \bibinfo
  {pages} {281} (\bibinfo {year} {2009})}\BibitemShut {NoStop}%
\bibitem [{\citenamefont {Kyrylyuk}\ and\ \citenamefont
  {Philipse}(2011)}]{Kyrylyuk2011}%
  \BibitemOpen
  \bibfield  {author} {\bibinfo {author} {\bibfnamefont {A.~V.}\ \bibnamefont
  {Kyrylyuk}}\ and\ \bibinfo {author} {\bibfnamefont {A.~P.}\ \bibnamefont
  {Philipse}},\ }\bibfield  {title} {\bibinfo {title} {Effect of particle shape
  on the random packing density of amorphous solids},\ }\href
  {https://doi.org/10.1002/pssa.201000361} {\bibfield  {journal} {\bibinfo
  {journal} {phys. status solidi (a)}\ }\textbf {\bibinfo {volume} {208}},\
  \bibinfo {pages} {2299} (\bibinfo {year} {2011})}\BibitemShut {NoStop}%
\bibitem [{\citenamefont {Kwan}\ and\ \citenamefont {Mora}(2001)}]{Kwan2001}%
  \BibitemOpen
  \bibfield  {author} {\bibinfo {author} {\bibfnamefont {A.~K.~H.}\
  \bibnamefont {Kwan}}\ and\ \bibinfo {author} {\bibfnamefont {C.~F.}\
  \bibnamefont {Mora}},\ }\bibfield  {title} {\bibinfo {title} {Effects of
  various shape parameters on packing of aggregate particles},\ }\href
  {https://doi.org/10.1680/macr.2001.53.2.91} {\bibfield  {journal} {\bibinfo
  {journal} {Mag. Concr. Res.}\ }\textbf {\bibinfo {volume} {53}},\ \bibinfo
  {pages} {91} (\bibinfo {year} {2001})}\BibitemShut {NoStop}%
\bibitem [{\citenamefont {Abreu}\ \emph {et~al.}(2003)\citenamefont {Abreu},
  \citenamefont {Tavares},\ and\ \citenamefont {Castier}}]{Abreu2003}%
  \BibitemOpen
  \bibfield  {author} {\bibinfo {author} {\bibfnamefont {C.~R.~A.}\
  \bibnamefont {Abreu}}, \bibinfo {author} {\bibfnamefont {F.~W.}\ \bibnamefont
  {Tavares}},\ and\ \bibinfo {author} {\bibfnamefont {M.}~\bibnamefont
  {Castier}},\ }\bibfield  {title} {\bibinfo {title} {Influence of particle
  shape on the packing and on the segregation of spherocylinders via {Monte}
  {Carlo} simulations},\ }\href {https://doi.org/10.1016/S0032-5910(03)00151-7}
  {\bibfield  {journal} {\bibinfo  {journal} {Powder Technol.}\ }\textbf
  {\bibinfo {volume} {134}},\ \bibinfo {pages} {167} (\bibinfo {year}
  {2003})}\BibitemShut {NoStop}%
\bibitem [{\citenamefont {Az\'ema}\ and\ \citenamefont
  {Radja\"{\i}}(2012)}]{Azema2012}%
  \BibitemOpen
  \bibfield  {author} {\bibinfo {author} {\bibfnamefont {E.}~\bibnamefont
  {Az\'ema}}\ and\ \bibinfo {author} {\bibfnamefont {F.}~\bibnamefont
  {Radja\"{\i}}},\ }\bibfield  {title} {\bibinfo {title} {Force chains and
  contact network topology in sheared packings of elongated particles},\ }\href
  {https://doi.org/10.1103/PhysRevE.85.031303} {\bibfield  {journal} {\bibinfo
  {journal} {Phys. Rev. E}\ }\textbf {\bibinfo {volume} {85}},\ \bibinfo
  {pages} {031303} (\bibinfo {year} {2012})}\BibitemShut {NoStop}%
\bibitem [{\citenamefont {Chen}\ and\ \citenamefont
  {Hlavacek}(1994)}]{Chen1994}%
  \BibitemOpen
  \bibfield  {author} {\bibinfo {author} {\bibfnamefont {V.}~\bibnamefont
  {Chen}}\ and\ \bibinfo {author} {\bibfnamefont {M.}~\bibnamefont
  {Hlavacek}},\ }\bibfield  {title} {\bibinfo {title} {Application of voronoi
  tessellation for modeling randomly packed hollow-fiber bundles},\ }\href
  {https://doi.org/10.1002/aic.690400405} {\bibfield  {journal} {\bibinfo
  {journal} {AIChE J.}\ }\textbf {\bibinfo {volume} {40}},\ \bibinfo {pages}
  {606} (\bibinfo {year} {1994})}\BibitemShut {NoStop}%
\bibitem [{\citenamefont {Bokobza}(2019)}]{Bokobza2019}%
  \BibitemOpen
  \bibfield  {author} {\bibinfo {author} {\bibfnamefont {L.}~\bibnamefont
  {Bokobza}},\ }\bibfield  {title} {\bibinfo {title} {Natural rubber
  nanocomposites: {A} review},\ }\href {https://doi.org/10.3390/nano9010012}
  {\bibfield  {journal} {\bibinfo  {journal} {NANOMATERIALS-BASEL}\ }\textbf
  {\bibinfo {volume} {9}},\ \bibinfo {pages} {12} (\bibinfo {year}
  {2019})}\BibitemShut {NoStop}%
\bibitem [{\citenamefont {Yang}\ \emph {et~al.}(2019)\citenamefont {Yang},
  \citenamefont {Zhou}, \citenamefont {Song},\ and\ \citenamefont
  {Chen}}]{Yang2019}%
  \BibitemOpen
  \bibfield  {author} {\bibinfo {author} {\bibfnamefont {L.}~\bibnamefont
  {Yang}}, \bibinfo {author} {\bibfnamefont {Z.}~\bibnamefont {Zhou}}, \bibinfo
  {author} {\bibfnamefont {J.}~\bibnamefont {Song}},\ and\ \bibinfo {author}
  {\bibfnamefont {X.}~\bibnamefont {Chen}},\ }\bibfield  {title} {\bibinfo
  {title} {Anisotropic nanomaterials for shape-dependent physicochemical and
  biomedical applications},\ }\href {https://doi.org/10.1039/C9CS00011A}
  {\bibfield  {journal} {\bibinfo  {journal} {Chem. Soc. Rev.}\ }\textbf
  {\bibinfo {volume} {48}},\ \bibinfo {pages} {5140} (\bibinfo {year}
  {2019})}\BibitemShut {NoStop}%
\bibitem [{\citenamefont {Pampaloni}\ \emph {et~al.}(2019)\citenamefont
  {Pampaloni}, \citenamefont {Giugliano}, \citenamefont {Scaini}, \citenamefont
  {Ballerini},\ and\ \citenamefont {Rauti}}]{Pampaloni2019}%
  \BibitemOpen
  \bibfield  {author} {\bibinfo {author} {\bibfnamefont {N.~P.}\ \bibnamefont
  {Pampaloni}}, \bibinfo {author} {\bibfnamefont {M.}~\bibnamefont
  {Giugliano}}, \bibinfo {author} {\bibfnamefont {D.}~\bibnamefont {Scaini}},
  \bibinfo {author} {\bibfnamefont {L.}~\bibnamefont {Ballerini}},\ and\
  \bibinfo {author} {\bibfnamefont {R.}~\bibnamefont {Rauti}},\ }\bibfield
  {title} {\bibinfo {title} {Advances in nano neuroscience: {From}
  nanomaterials to nanotools},\ }\href
  {https://doi.org/10.3389/fnins.2018.00953} {\bibfield  {journal} {\bibinfo
  {journal} {Front. Neurosci.}\ }\textbf {\bibinfo {volume} {12}},\ \bibinfo
  {pages} {953} (\bibinfo {year} {2019})}\BibitemShut {NoStop}%
\bibitem [{\citenamefont {Lebovka}\ \emph
  {et~al.}(2019{\natexlab{a}})\citenamefont {Lebovka}, \citenamefont
  {Lisetski},\ and\ \citenamefont {Bulavin}}]{Lebovka2019Springer}%
  \BibitemOpen
  \bibfield  {author} {\bibinfo {author} {\bibfnamefont {N.}~\bibnamefont
  {Lebovka}}, \bibinfo {author} {\bibfnamefont {L.}~\bibnamefont {Lisetski}},\
  and\ \bibinfo {author} {\bibfnamefont {L.}~\bibnamefont {Bulavin}},\
  }\bibfield  {title} {\bibinfo {title} {Organization of nano-disks of
  laponite\textsuperscript{$\circledR$} in soft colloidal systems},\ }in\ \href
  {https://doi.org/10.1007/978-3-030-21755-6_6} {\emph {\bibinfo {booktitle}
  {Modern Problems of the Physics of Liquid Systems}}}\ (\bibinfo  {publisher}
  {Springer},\ \bibinfo {year} {2019})\ pp.\ \bibinfo {pages}
  {137--164}\BibitemShut {NoStop}%
\bibitem [{\citenamefont {Wang}\ \emph {et~al.}(2020)\citenamefont {Wang},
  \citenamefont {Liu}, \citenamefont {Qi}, \citenamefont {Al-Tabbaa},\ and\
  \citenamefont {Wang}}]{Wang2020}%
  \BibitemOpen
  \bibfield  {author} {\bibinfo {author} {\bibfnamefont {M.}~\bibnamefont
  {Wang}}, \bibinfo {author} {\bibfnamefont {Y.}~\bibnamefont {Liu}}, \bibinfo
  {author} {\bibfnamefont {B.}~\bibnamefont {Qi}}, \bibinfo {author}
  {\bibfnamefont {A.}~\bibnamefont {Al-Tabbaa}},\ and\ \bibinfo {author}
  {\bibfnamefont {W.}~\bibnamefont {Wang}},\ }\bibfield  {title} {\bibinfo
  {title} {Percolation and conductivity development of the rod networks within
  randomly packed porous media},\ }\href
  {https://doi.org/10.1016/j.compositesb.2020.107837} {\bibfield  {journal}
  {\bibinfo  {journal} {Compos. Part B Eng.}\ }\textbf {\bibinfo {volume}
  {187}},\ \bibinfo {pages} {107837} (\bibinfo {year} {2020})}\BibitemShut
  {NoStop}%
\bibitem [{\citenamefont {Hirotani}\ and\ \citenamefont
  {Ohno}(2019)}]{Hirotani2019}%
  \BibitemOpen
  \bibfield  {author} {\bibinfo {author} {\bibfnamefont {J.}~\bibnamefont
  {Hirotani}}\ and\ \bibinfo {author} {\bibfnamefont {Y.}~\bibnamefont
  {Ohno}},\ }\bibinfo {title} {Carbon nanotube thin films for high-performance
  flexible electronics applications},\ in\ \href
  {https://doi.org/10.1007/978-3-030-12700-8_9} {\emph {\bibinfo {booktitle}
  {Single-Walled Carbon Nanotubes: Preparation, Properties and
  Applications}}},\ \bibinfo {editor} {edited by\ \bibinfo {editor}
  {\bibfnamefont {Y.}~\bibnamefont {Li}}\ and\ \bibinfo {editor} {\bibfnamefont
  {S.}~\bibnamefont {Maruyama}}}\ (\bibinfo  {publisher} {Springer
  International Publishing},\ \bibinfo {address} {Cham},\ \bibinfo {year}
  {2019})\ pp.\ \bibinfo {pages} {257--270}\BibitemShut {NoStop}%
\bibitem [{\citenamefont {Tiginyanu}\ \emph {et~al.}(2019)\citenamefont
  {Tiginyanu}, \citenamefont {Topala},\ and\ \citenamefont
  {Ursaki}}]{Tiginyanu2019}%
  \BibitemOpen
  \bibinfo {editor} {\bibfnamefont {I.}~\bibnamefont {Tiginyanu}}, \bibinfo
  {editor} {\bibfnamefont {P.}~\bibnamefont {Topala}},\ and\ \bibinfo {editor}
  {\bibfnamefont {V.}~\bibnamefont {Ursaki}},\ eds.,\ \href
  {https://doi.org/10.1007/978-3-319-30198-3} {\emph {\bibinfo {title}
  {Nanostructures and thin films for multifunctional applications. Technology,
  Properties and Devices}}}\ (\bibinfo  {publisher} {Springer Nature},\
  \bibinfo {year} {2019})\BibitemShut {NoStop}%
\bibitem [{\citenamefont {Lebovka}\ and\ \citenamefont
  {Tarasevich}(2020)}]{Lebovka2020}%
  \BibitemOpen
  \bibfield  {author} {\bibinfo {author} {\bibfnamefont {N.~I.}\ \bibnamefont
  {Lebovka}}\ and\ \bibinfo {author} {\bibfnamefont {Y.~Y.}\ \bibnamefont
  {Tarasevich}},\ }\bibfield  {title} {\bibinfo {title} {Two-dimensional
  systems of elongated particles: {From} diluted to dense},\ }in\ \href
  {https://doi.org/10.1142/11711} {\emph {\bibinfo {booktitle} {Order, Disorder
  and Criticality. Advanced Problems of Phase Transition Theory}}},\
  Vol.~\bibinfo {volume} {6},\ \bibinfo {editor} {edited by\ \bibinfo {editor}
  {\bibfnamefont {Y.}~\bibnamefont {Holovatch}}}\ (\bibinfo  {publisher} {World
  Scientific, Singapore},\ \bibinfo {year} {2020})\ pp.\ \bibinfo {pages}
  {153--183},\ \bibinfo {note} {arXiv:1912.09799}\BibitemShut {NoStop}%
\bibitem [{\citenamefont {Basurto}\ \emph {et~al.}(2020)\citenamefont
  {Basurto}, \citenamefont {Gurin}, \citenamefont {Varga},\ and\ \citenamefont
  {Odriozola}}]{Basurto2020}%
  \BibitemOpen
  \bibfield  {author} {\bibinfo {author} {\bibfnamefont {E.}~\bibnamefont
  {Basurto}}, \bibinfo {author} {\bibfnamefont {P.}~\bibnamefont {Gurin}},
  \bibinfo {author} {\bibfnamefont {S.}~\bibnamefont {Varga}},\ and\ \bibinfo
  {author} {\bibfnamefont {G.}~\bibnamefont {Odriozola}},\ }\bibfield  {title}
  {\bibinfo {title} {Ordering, clustering, and wetting of hard rods in extreme
  confinement},\ }\href {https://doi.org/10.1103/PhysRevResearch.2.013356}
  {\bibfield  {journal} {\bibinfo  {journal} {Phys. Rev. Research}\ }\textbf
  {\bibinfo {volume} {2}},\ \bibinfo {pages} {013356} (\bibinfo {year}
  {2020})}\BibitemShut {NoStop}%
\bibitem [{\citenamefont {Frenkel}\ and\ \citenamefont
  {Eppenga}(1985)}]{Frenkel1985}%
  \BibitemOpen
  \bibfield  {author} {\bibinfo {author} {\bibfnamefont {D.}~\bibnamefont
  {Frenkel}}\ and\ \bibinfo {author} {\bibfnamefont {R.}~\bibnamefont
  {Eppenga}},\ }\bibfield  {title} {\bibinfo {title} {Evidence for algebraic
  orientational order in a two-dimensional hard-core nematic},\ }\href
  {https://doi.org/10.1103/PhysRevA.31.1776} {\bibfield  {journal} {\bibinfo
  {journal} {Phys. Rev. A}\ }\textbf {\bibinfo {volume} {31}},\ \bibinfo
  {pages} {1776} (\bibinfo {year} {1985})}\BibitemShut {NoStop}%
\bibitem [{\citenamefont {Bates}\ and\ \citenamefont
  {Frenkel}(2000)}]{Bates2000}%
  \BibitemOpen
  \bibfield  {author} {\bibinfo {author} {\bibfnamefont {M.~A.}\ \bibnamefont
  {Bates}}\ and\ \bibinfo {author} {\bibfnamefont {D.}~\bibnamefont
  {Frenkel}},\ }\bibfield  {title} {\bibinfo {title} {Phase behavior of
  two-dimensional hard rod fluids},\ }\href {https://doi.org/10.1063/1.481637}
  {\bibfield  {journal} {\bibinfo  {journal} {J. Chem. Phys.}\ }\textbf
  {\bibinfo {volume} {112}},\ \bibinfo {pages} {10034} (\bibinfo {year}
  {2000})}\BibitemShut {NoStop}%
\bibitem [{\citenamefont {Evans}(1993)}]{Evans1993}%
  \BibitemOpen
  \bibfield  {author} {\bibinfo {author} {\bibfnamefont {J.~W.}\ \bibnamefont
  {Evans}},\ }\bibfield  {title} {\bibinfo {title} {Random and cooperative
  sequential adsorption},\ }\href {https://doi.org/10.1103/RevModPhys.65.1281}
  {\bibfield  {journal} {\bibinfo  {journal} {Rev. Mod. Phys.}\ }\textbf
  {\bibinfo {volume} {65}},\ \bibinfo {pages} {1281} (\bibinfo {year}
  {1993})}\BibitemShut {NoStop}%
\bibitem [{\citenamefont {Adamczyk}(2017)}]{Adamczyk2017}%
  \BibitemOpen
  \bibfield  {author} {\bibinfo {author} {\bibfnamefont {Z.}~\bibnamefont
  {Adamczyk}},\ }\href@noop {} {\emph {\bibinfo {title} {Particles at
  interfaces}}},\ \bibinfo {edition} {2nd}\ ed.,\ \bibinfo {series}
  {Interactions, deposition, structure}, Vol.~\bibinfo {volume} {20}\ (\bibinfo
   {publisher} {Elsevier},\ \bibinfo {address} {London, UK},\ \bibinfo {year}
  {2017})\BibitemShut {NoStop}%
\bibitem [{\citenamefont {Finegold}\ and\ \citenamefont
  {Donnell}(1979)}]{Finegold1979Nat}%
  \BibitemOpen
  \bibfield  {author} {\bibinfo {author} {\bibfnamefont {L.}~\bibnamefont
  {Finegold}}\ and\ \bibinfo {author} {\bibfnamefont {J.~T.}\ \bibnamefont
  {Donnell}},\ }\bibfield  {title} {\bibinfo {title} {Maximum density of random
  placing of membrane particles},\ }\href {https://doi.org/10.1038/278443a0}
  {\bibfield  {journal} {\bibinfo  {journal} {Nature}\ }\textbf {\bibinfo
  {volume} {278}},\ \bibinfo {pages} {443} (\bibinfo {year}
  {1979})}\BibitemShut {NoStop}%
\bibitem [{\citenamefont {Feder}(1980)}]{Feder1980}%
  \BibitemOpen
  \bibfield  {author} {\bibinfo {author} {\bibfnamefont {J.}~\bibnamefont
  {Feder}},\ }\bibfield  {title} {\bibinfo {title} {Random sequential
  adsorption},\ }\href {https://doi.org/10.1016/0022-5193(80)90358-6}
  {\bibfield  {journal} {\bibinfo  {journal} {J. Theor. Biol.}\ }\textbf
  {\bibinfo {volume} {87}},\ \bibinfo {pages} {237} (\bibinfo {year}
  {1980})}\BibitemShut {NoStop}%
\bibitem [{\citenamefont {Sherwood}(1990)}]{Sherwood1990}%
  \BibitemOpen
  \bibfield  {author} {\bibinfo {author} {\bibfnamefont {J.~D.}\ \bibnamefont
  {Sherwood}},\ }\bibfield  {title} {\bibinfo {title} {Random sequential
  adsorption of lines and ellipses},\ }\href
  {https://doi.org/10.1088/0305-4470/23/13/021} {\bibfield  {journal} {\bibinfo
   {journal} {J. Phys. A: Math. Gen.}\ }\textbf {\bibinfo {volume} {23}},\
  \bibinfo {pages} {2827} (\bibinfo {year} {1990})}\BibitemShut {NoStop}%
\bibitem [{\citenamefont {Viot}\ and\ \citenamefont {Tarjus}(1990)}]{Viot1990}%
  \BibitemOpen
  \bibfield  {author} {\bibinfo {author} {\bibfnamefont {P.}~\bibnamefont
  {Viot}}\ and\ \bibinfo {author} {\bibfnamefont {G.}~\bibnamefont {Tarjus}},\
  }\bibfield  {title} {\bibinfo {title} {Random sequential addition of
  unoriented squares: {Breakdown} of {Swendsen's} conjecture},\ }\href
  {https://doi.org/10.1209/0295-5075/13/4/002} {\bibfield  {journal} {\bibinfo
  {journal} {EPL (Europhysics Letters)}\ }\textbf {\bibinfo {volume} {13}},\
  \bibinfo {pages} {295} (\bibinfo {year} {1990})}\BibitemShut {NoStop}%
\bibitem [{\citenamefont {Vigil}\ and\ \citenamefont {Ziff}(1989)}]{Vigil1989}%
  \BibitemOpen
  \bibfield  {author} {\bibinfo {author} {\bibfnamefont {R.~D.}\ \bibnamefont
  {Vigil}}\ and\ \bibinfo {author} {\bibfnamefont {R.~M.}\ \bibnamefont
  {Ziff}},\ }\bibfield  {title} {\bibinfo {title} {Random sequential adsorption
  of unoriented rectangles onto a plane},\ }\href
  {https://doi.org/10.1063/1.457021} {\bibfield  {journal} {\bibinfo  {journal}
  {J. Chem. Phys.}\ }\textbf {\bibinfo {volume} {91}},\ \bibinfo {pages} {2599}
  (\bibinfo {year} {1989})}\BibitemShut {NoStop}%
\bibitem [{\citenamefont {Vigil}\ and\ \citenamefont {Ziff}(1990)}]{Vigil1990}%
  \BibitemOpen
  \bibfield  {author} {\bibinfo {author} {\bibfnamefont {R.~D.}\ \bibnamefont
  {Vigil}}\ and\ \bibinfo {author} {\bibfnamefont {R.~M.}\ \bibnamefont
  {Ziff}},\ }\bibfield  {title} {\bibinfo {title} {Kinetics of random
  sequential adsorption of rectangles and line segments},\ }\href
  {https://doi.org/10.1063/1.459307} {\bibfield  {journal} {\bibinfo  {journal}
  {J. Chem. Phys.}\ }\textbf {\bibinfo {volume} {93}},\ \bibinfo {pages} {8270}
  (\bibinfo {year} {1990})}\BibitemShut {NoStop}%
\bibitem [{\citenamefont {Haiduk}\ \emph {et~al.}(2018)\citenamefont {Haiduk},
  \citenamefont {Kubala},\ and\ \citenamefont {Cie\'{s}la}}]{Haiduk2018}%
  \BibitemOpen
  \bibfield  {author} {\bibinfo {author} {\bibfnamefont {K.}~\bibnamefont
  {Haiduk}}, \bibinfo {author} {\bibfnamefont {P.}~\bibnamefont {Kubala}},\
  and\ \bibinfo {author} {\bibfnamefont {M.}~\bibnamefont {Cie\'{s}la}},\
  }\bibfield  {title} {\bibinfo {title} {Saturated packings of convex
  anisotropic objects under random sequential adsorption protocol},\ }\href
  {https://doi.org/10.1103/PhysRevE.98.063309} {\bibfield  {journal} {\bibinfo
  {journal} {Phys. Rev. E}\ }\textbf {\bibinfo {volume} {98}},\ \bibinfo
  {pages} {063309} (\bibinfo {year} {2018})}\BibitemShut {NoStop}%
\bibitem [{\citenamefont {Cie{\'s}la}\ and\ \citenamefont
  {Barbasz}(2014)}]{Ciesla2014}%
  \BibitemOpen
  \bibfield  {author} {\bibinfo {author} {\bibfnamefont {M.}~\bibnamefont
  {Cie{\'s}la}}\ and\ \bibinfo {author} {\bibfnamefont {J.}~\bibnamefont
  {Barbasz}},\ }\bibfield  {title} {\bibinfo {title} {Random packing of regular
  polygons and star polygons on a flat two-dimensional surface},\ }\href
  {https://doi.org/10.1103/PhysRevE.90.022402} {\bibfield  {journal} {\bibinfo
  {journal} {Phys. Rev. E}\ }\textbf {\bibinfo {volume} {90}},\ \bibinfo
  {pages} {022402} (\bibinfo {year} {2014})}\BibitemShut {NoStop}%
\bibitem [{\citenamefont {Zhang}(2018)}]{Zhang2018}%
  \BibitemOpen
  \bibfield  {author} {\bibinfo {author} {\bibfnamefont {G.}~\bibnamefont
  {Zhang}},\ }\bibfield  {title} {\bibinfo {title} {Precise algorithm to
  generate random sequential adsorption of hard polygons at saturation},\
  }\href {https://doi.org/10.1103/PhysRevE.97.043311} {\bibfield  {journal}
  {\bibinfo  {journal} {Phys. Rev. E}\ }\textbf {\bibinfo {volume} {97}},\
  \bibinfo {pages} {043311} (\bibinfo {year} {2018})}\BibitemShut {NoStop}%
\bibitem [{\citenamefont {Cie{\'s}la}\ and\ \citenamefont
  {Barbasz}(2013)}]{Ciesla2013}%
  \BibitemOpen
  \bibfield  {author} {\bibinfo {author} {\bibfnamefont {M.}~\bibnamefont
  {Cie{\'s}la}}\ and\ \bibinfo {author} {\bibfnamefont {J.}~\bibnamefont
  {Barbasz}},\ }\bibfield  {title} {\bibinfo {title} {Modelling of interacting
  dimer adsorption},\ }\href {https://doi.org/10.1016/j.susc.2013.02.013}
  {\bibfield  {journal} {\bibinfo  {journal} {Surf. Sci.}\ }\textbf {\bibinfo
  {volume} {612}},\ \bibinfo {pages} {24} (\bibinfo {year} {2013})}\BibitemShut
  {NoStop}%
\bibitem [{\citenamefont {Cie{\'s}la}(2013)}]{Ciesla2013a}%
  \BibitemOpen
  \bibfield  {author} {\bibinfo {author} {\bibfnamefont {M.}~\bibnamefont
  {Cie{\'s}la}},\ }\bibfield  {title} {\bibinfo {title} {Continuum random
  sequential adsorption of polymer on a flat and homogeneous surface},\ }\href
  {https://doi.org/10.1103/PhysRevE.87.052401} {\bibfield  {journal} {\bibinfo
  {journal} {Phys. Rev. E}\ }\textbf {\bibinfo {volume} {87}},\ \bibinfo
  {pages} {052401} (\bibinfo {year} {2013})}\BibitemShut {NoStop}%
\bibitem [{\citenamefont {Cie{\'s}la}\ \emph {et~al.}(2015)\citenamefont
  {Cie{\'s}la}, \citenamefont {Paja},\ and\ \citenamefont {Ziff}}]{Ciesla2015}%
  \BibitemOpen
  \bibfield  {author} {\bibinfo {author} {\bibfnamefont {M.}~\bibnamefont
  {Cie{\'s}la}}, \bibinfo {author} {\bibfnamefont {G.}~\bibnamefont {Paja}},\
  and\ \bibinfo {author} {\bibfnamefont {R.~M.}\ \bibnamefont {Ziff}},\
  }\bibfield  {title} {\bibinfo {title} {Shapes for maximal coverage for
  two-dimensional random sequential adsorption},\ }\href
  {https://doi.org/10.1039/C5CP03873A} {\bibfield  {journal} {\bibinfo
  {journal} {Phys. Chem. Chem. Phys.}\ }\textbf {\bibinfo {volume} {17}},\
  \bibinfo {pages} {24376} (\bibinfo {year} {2015})}\BibitemShut {NoStop}%
\bibitem [{\citenamefont {Viot}\ \emph {et~al.}(1992)\citenamefont {Viot},
  \citenamefont {Tarjus}, \citenamefont {Ricci},\ and\ \citenamefont
  {Talbot}}]{Viot1992}%
  \BibitemOpen
  \bibfield  {author} {\bibinfo {author} {\bibfnamefont {P.}~\bibnamefont
  {Viot}}, \bibinfo {author} {\bibfnamefont {G.}~\bibnamefont {Tarjus}},
  \bibinfo {author} {\bibfnamefont {S.~M.}\ \bibnamefont {Ricci}},\ and\
  \bibinfo {author} {\bibfnamefont {J.}~\bibnamefont {Talbot}},\ }\bibfield
  {title} {\bibinfo {title} {Random sequential adsorption of anisotropic
  particles. {I.} {Jamming} limit and asymptotic behavior},\ }\href
  {https://doi.org/10.1063/1.463820} {\bibfield  {journal} {\bibinfo  {journal}
  {J. Chem. Phys.}\ }\textbf {\bibinfo {volume} {97}},\ \bibinfo {pages} {5212}
  (\bibinfo {year} {1992})}\BibitemShut {NoStop}%
\bibitem [{\citenamefont {Chaikin}\ \emph {et~al.}(2006)\citenamefont
  {Chaikin}, \citenamefont {Donev}, \citenamefont {Man}, \citenamefont
  {Stillinger},\ and\ \citenamefont {Torquato}}]{Chaikin2006}%
  \BibitemOpen
  \bibfield  {author} {\bibinfo {author} {\bibfnamefont {P.~M.}\ \bibnamefont
  {Chaikin}}, \bibinfo {author} {\bibfnamefont {A.}~\bibnamefont {Donev}},
  \bibinfo {author} {\bibfnamefont {W.}~\bibnamefont {Man}}, \bibinfo {author}
  {\bibfnamefont {F.~H.}\ \bibnamefont {Stillinger}},\ and\ \bibinfo {author}
  {\bibfnamefont {S.}~\bibnamefont {Torquato}},\ }\bibfield  {title} {\bibinfo
  {title} {Some observations on the random packing of hard ellipsoids},\ }\href
  {https://doi.org/10.1021/ie060032g} {\bibfield  {journal} {\bibinfo
  {journal} {Ind. Eng. Chem. Res.}\ }\textbf {\bibinfo {volume} {45}},\
  \bibinfo {pages} {6960} (\bibinfo {year} {2006})}\BibitemShut {NoStop}%
\bibitem [{\citenamefont {Baule}(2017)}]{Baule2017}%
  \BibitemOpen
  \bibfield  {author} {\bibinfo {author} {\bibfnamefont {A.}~\bibnamefont
  {Baule}},\ }\bibfield  {title} {\bibinfo {title} {Shape universality classes
  in the random sequential adsorption of nonspherical particles},\ }\href
  {https://doi.org/10.1103/PhysRevLett.119.028003} {\bibfield  {journal}
  {\bibinfo  {journal} {Phys. Rev. Lett.}\ }\textbf {\bibinfo {volume} {119}},\
  \bibinfo {pages} {028003} (\bibinfo {year} {2017})}\BibitemShut {NoStop}%
\bibitem [{\citenamefont {Cie{\'s}la}\ \emph {et~al.}(2020)\citenamefont
  {Cie{\'s}la}, \citenamefont {Kozubek}, \citenamefont {Kubala},\ and\
  \citenamefont {Baule}}]{Ciesla2020}%
  \BibitemOpen
  \bibfield  {author} {\bibinfo {author} {\bibfnamefont {M.}~\bibnamefont
  {Cie{\'s}la}}, \bibinfo {author} {\bibfnamefont {K.}~\bibnamefont {Kozubek}},
  \bibinfo {author} {\bibfnamefont {P.}~\bibnamefont {Kubala}},\ and\ \bibinfo
  {author} {\bibfnamefont {A.}~\bibnamefont {Baule}},\ }\bibfield  {title}
  {\bibinfo {title} {Kinetics of random sequential adsorption of
  two-dimensional shapes on a one-dimensional line},\ }\href
  {https://doi.org/10.1103/PhysRevE.101.042901} {\bibfield  {journal} {\bibinfo
   {journal} {Phys. Rev. E}\ }\textbf {\bibinfo {volume} {101}},\ \bibinfo
  {pages} {042901} (\bibinfo {year} {2020})}\BibitemShut {NoStop}%
\bibitem [{\citenamefont {Lebovka}\ \emph {et~al.}(2020)\citenamefont
  {Lebovka}, \citenamefont {Tatochenko}, \citenamefont {Vygornitskii},\ and\
  \citenamefont {Tarasevich}}]{Lebovka2020a}%
  \BibitemOpen
  \bibfield  {author} {\bibinfo {author} {\bibfnamefont {N.~I.}\ \bibnamefont
  {Lebovka}}, \bibinfo {author} {\bibfnamefont {M.~O.}\ \bibnamefont
  {Tatochenko}}, \bibinfo {author} {\bibfnamefont {N.~V.}\ \bibnamefont
  {Vygornitskii}},\ and\ \bibinfo {author} {\bibfnamefont {Y.~Y.}\ \bibnamefont
  {Tarasevich}},\ }\href@noop {} {\bibinfo {title} {``{Paris} car parking
  problem'' for partially ordered discorectangles on a line}} (\bibinfo {year}
  {2020}),\ \Eprint {https://arxiv.org/abs/2005.04206} {arXiv:2005.04206
  [cond-mat.dis-nn]} \BibitemShut {NoStop}%
\bibitem [{\citenamefont {Brosilow}\ \emph {et~al.}(1991)\citenamefont
  {Brosilow}, \citenamefont {Ziff},\ and\ \citenamefont
  {Vigil}}]{Brosilow1991}%
  \BibitemOpen
  \bibfield  {author} {\bibinfo {author} {\bibfnamefont {B.~J.}\ \bibnamefont
  {Brosilow}}, \bibinfo {author} {\bibfnamefont {R.~M.}\ \bibnamefont {Ziff}},\
  and\ \bibinfo {author} {\bibfnamefont {R.~D.}\ \bibnamefont {Vigil}},\
  }\bibfield  {title} {\bibinfo {title} {Random sequential adsorption of
  parallel squares},\ }\href {https://doi.org/10.1103/PhysRevA.43.631}
  {\bibfield  {journal} {\bibinfo  {journal} {Phys. Rev. A}\ }\textbf {\bibinfo
  {volume} {43}},\ \bibinfo {pages} {631} (\bibinfo {year} {1991})}\BibitemShut
  {NoStop}%
\bibitem [{\citenamefont {Gromenko}\ and\ \citenamefont
  {Privman}(2009)}]{Gromenko2009}%
  \BibitemOpen
  \bibfield  {author} {\bibinfo {author} {\bibfnamefont {O.}~\bibnamefont
  {Gromenko}}\ and\ \bibinfo {author} {\bibfnamefont {V.}~\bibnamefont
  {Privman}},\ }\bibfield  {title} {\bibinfo {title} {Random sequential
  adsorption of oriented superdisks},\ }\href
  {https://doi.org/10.1103/PhysRevE.79.042103} {\bibfield  {journal} {\bibinfo
  {journal} {Phys. Rev. E}\ }\textbf {\bibinfo {volume} {79}},\ \bibinfo
  {pages} {042103} (\bibinfo {year} {2009})}\BibitemShut {NoStop}%
\bibitem [{\citenamefont {Lebovka}\ \emph
  {et~al.}(2019{\natexlab{b}})\citenamefont {Lebovka}, \citenamefont
  {Vygornitskii},\ and\ \citenamefont {Tarasevich}}]{Lebovka2019}%
  \BibitemOpen
  \bibfield  {author} {\bibinfo {author} {\bibfnamefont {N.~I.}\ \bibnamefont
  {Lebovka}}, \bibinfo {author} {\bibfnamefont {N.~V.}\ \bibnamefont
  {Vygornitskii}},\ and\ \bibinfo {author} {\bibfnamefont {Y.~Y.}\ \bibnamefont
  {Tarasevich}},\ }\bibfield  {title} {\bibinfo {title} {Relaxation in
  two-dimensional suspensions of rods as driven by {Brownian} diffusion},\
  }\href {https://doi.org/10.1103/PhysRevE.100.042139} {\bibfield  {journal}
  {\bibinfo  {journal} {Phys. Rev. E}\ }\textbf {\bibinfo {volume} {100}},\
  \bibinfo {pages} {042139} (\bibinfo {year} {2019}{\natexlab{b}})}\BibitemShut
  {NoStop}%
\bibitem [{\citenamefont {Ackermann}\ \emph {et~al.}(2016)\citenamefont
  {Ackermann}, \citenamefont {Neuhaus},\ and\ \citenamefont
  {Roth}}]{Ackermann2016}%
  \BibitemOpen
  \bibfield  {author} {\bibinfo {author} {\bibfnamefont {T.}~\bibnamefont
  {Ackermann}}, \bibinfo {author} {\bibfnamefont {R.}~\bibnamefont {Neuhaus}},\
  and\ \bibinfo {author} {\bibfnamefont {S.}~\bibnamefont {Roth}},\ }\bibfield
  {title} {\bibinfo {title} {The effect of rod orientation on electrical
  anisotropy in silver nanowire networks for ultra-transparent electrodes},\
  }\href {https://doi.org/10.1038/srep34289} {\bibfield  {journal} {\bibinfo
  {journal} {Sci. Rep.}\ }\textbf {\bibinfo {volume} {6}},\ \bibinfo {pages}
  {34289} (\bibinfo {year} {2016})}\BibitemShut {NoStop}%
\bibitem [{\citenamefont {Wu}\ \emph {et~al.}(2017)\citenamefont {Wu},
  \citenamefont {Jiang}, \citenamefont {Zan}, \citenamefont {Lin},\ and\
  \citenamefont {Wang}}]{Wu2017}%
  \BibitemOpen
  \bibfield  {author} {\bibinfo {author} {\bibfnamefont {Y.}~\bibnamefont
  {Wu}}, \bibinfo {author} {\bibfnamefont {Z.}~\bibnamefont {Jiang}}, \bibinfo
  {author} {\bibfnamefont {X.}~\bibnamefont {Zan}}, \bibinfo {author}
  {\bibfnamefont {Y.}~\bibnamefont {Lin}},\ and\ \bibinfo {author}
  {\bibfnamefont {Q.}~\bibnamefont {Wang}},\ }\bibfield  {title} {\bibinfo
  {title} {Shear flow induced long-range ordering of rod-like viral
  nanoparticles within hydrogel},\ }\href {https://doi.org/10.1038/srep34289}
  {\bibfield  {journal} {\bibinfo  {journal} {Colloids Surf., B}\ }\textbf
  {\bibinfo {volume} {158}},\ \bibinfo {pages} {620} (\bibinfo {year}
  {2017})}\BibitemShut {NoStop}%
\bibitem [{\citenamefont {Shaver}\ \emph {et~al.}(2009)\citenamefont {Shaver},
  \citenamefont {Parra-Vasquez}, \citenamefont {Hansel}, \citenamefont
  {Portugall}, \citenamefont {Mielke}, \citenamefont {von Ortenberg},
  \citenamefont {Hauge}, \citenamefont {Pasquali},\ and\ \citenamefont
  {Kono}}]{Shaver2009}%
  \BibitemOpen
  \bibfield  {author} {\bibinfo {author} {\bibfnamefont {J.}~\bibnamefont
  {Shaver}}, \bibinfo {author} {\bibfnamefont {A.~N.~G.}\ \bibnamefont
  {Parra-Vasquez}}, \bibinfo {author} {\bibfnamefont {S.}~\bibnamefont
  {Hansel}}, \bibinfo {author} {\bibfnamefont {O.}~\bibnamefont {Portugall}},
  \bibinfo {author} {\bibfnamefont {C.~H.}\ \bibnamefont {Mielke}}, \bibinfo
  {author} {\bibfnamefont {M.}~\bibnamefont {von Ortenberg}}, \bibinfo {author}
  {\bibfnamefont {R.~H.}\ \bibnamefont {Hauge}}, \bibinfo {author}
  {\bibfnamefont {M.}~\bibnamefont {Pasquali}},\ and\ \bibinfo {author}
  {\bibfnamefont {J.}~\bibnamefont {Kono}},\ }\bibfield  {title} {\bibinfo
  {title} {Alignment dynamics of single-walled carbon nanotubes in pulsed
  ultrahigh magnetic fields},\ }\href {https://doi.org/10.1021/nn800519n}
  {\bibfield  {journal} {\bibinfo  {journal} {ACS Nano}\ }\textbf {\bibinfo
  {volume} {3}},\ \bibinfo {pages} {131} (\bibinfo {year} {2009})}\BibitemShut
  {NoStop}%
\bibitem [{\citenamefont {Mohammadimasoudi}\ \emph {et~al.}(2016)\citenamefont
  {Mohammadimasoudi}, \citenamefont {Hens},\ and\ \citenamefont
  {Neyts}}]{Mohammadimasoudi2016}%
  \BibitemOpen
  \bibfield  {author} {\bibinfo {author} {\bibfnamefont {M.}~\bibnamefont
  {Mohammadimasoudi}}, \bibinfo {author} {\bibfnamefont {Z.}~\bibnamefont
  {Hens}},\ and\ \bibinfo {author} {\bibfnamefont {K.}~\bibnamefont {Neyts}},\
  }\bibfield  {title} {\bibinfo {title} {Full alignment of dispersed colloidal
  nanorods by alternating electric fields},\ }\href
  {https://doi.org/10.1039/C6RA02620F} {\bibfield  {journal} {\bibinfo
  {journal} {RSC Adv.}\ }\textbf {\bibinfo {volume} {6}},\ \bibinfo {pages}
  {55736} (\bibinfo {year} {2016})}\BibitemShut {NoStop}%
\bibitem [{\citenamefont {van~der Zande}\ \emph {et~al.}(1999)\citenamefont
  {van~der Zande}, \citenamefont {Koper},\ and\ \citenamefont
  {Lekkerkerker}}]{Zande1999}%
  \BibitemOpen
  \bibfield  {author} {\bibinfo {author} {\bibfnamefont {B.~M.~I.}\
  \bibnamefont {van~der Zande}}, \bibinfo {author} {\bibfnamefont {G.~J.~M.}\
  \bibnamefont {Koper}},\ and\ \bibinfo {author} {\bibfnamefont {H.~N.~W.}\
  \bibnamefont {Lekkerkerker}},\ }\bibfield  {title} {\bibinfo {title}
  {Alignment of rod-shaped gold particles by electric fields},\ }\href
  {https://doi.org/10.1021/jp984737a} {\bibfield  {journal} {\bibinfo
  {journal} {J. Phys. Chem. B}\ }\textbf {\bibinfo {volume} {103}},\ \bibinfo
  {pages} {5754} (\bibinfo {year} {1999})}\BibitemShut {NoStop}%
\bibitem [{\citenamefont {Vega}\ and\ \citenamefont {Lago}(1994)}]{Vega1994}%
  \BibitemOpen
  \bibfield  {author} {\bibinfo {author} {\bibfnamefont {C.}~\bibnamefont
  {Vega}}\ and\ \bibinfo {author} {\bibfnamefont {S.}~\bibnamefont {Lago}},\
  }\bibfield  {title} {\bibinfo {title} {A fast algorithm to evaluate the
  shortest distance between rods},\ }\href
  {https://doi.org/10.1016/0097-8485(94)80023-5} {\bibfield  {journal}
  {\bibinfo  {journal} {Comput. Chem.}\ }\textbf {\bibinfo {volume} {18}},\
  \bibinfo {pages} {55} (\bibinfo {year} {1994})}\BibitemShut {NoStop}%
\bibitem [{\citenamefont {Pournin}\ \emph {et~al.}(2005)\citenamefont
  {Pournin}, \citenamefont {Weber}, \citenamefont {Tsukahara}, \citenamefont
  {Ferrez}, \citenamefont {Ramaioli},\ and\ \citenamefont
  {Liebling}}]{Pournin2005}%
  \BibitemOpen
  \bibfield  {author} {\bibinfo {author} {\bibfnamefont {L.}~\bibnamefont
  {Pournin}}, \bibinfo {author} {\bibfnamefont {M.}~\bibnamefont {Weber}},
  \bibinfo {author} {\bibfnamefont {M.}~\bibnamefont {Tsukahara}}, \bibinfo
  {author} {\bibfnamefont {J.-A.}\ \bibnamefont {Ferrez}}, \bibinfo {author}
  {\bibfnamefont {M.}~\bibnamefont {Ramaioli}},\ and\ \bibinfo {author}
  {\bibfnamefont {T.~M.}\ \bibnamefont {Liebling}},\ }\bibfield  {title}
  {\bibinfo {title} {Three-dimensional distinct element simulation of
  spherocylinder crystallization},\ }\href
  {https://doi.org/10.1007/s10035-004-0188-4} {\bibfield  {journal} {\bibinfo
  {journal} {Granul. Matter}\ }\textbf {\bibinfo {volume} {7}},\ \bibinfo
  {pages} {119} (\bibinfo {year} {2005})}\BibitemShut {NoStop}%
\bibitem [{\citenamefont {Mahajan}\ \emph {et~al.}(2018)\citenamefont
  {Mahajan}, \citenamefont {Nijssen}, \citenamefont {Kuipers},\ and\
  \citenamefont {Padding}}]{Mahajan2018}%
  \BibitemOpen
  \bibfield  {author} {\bibinfo {author} {\bibfnamefont {V.~V.}\ \bibnamefont
  {Mahajan}}, \bibinfo {author} {\bibfnamefont {T.~M.~J.}\ \bibnamefont
  {Nijssen}}, \bibinfo {author} {\bibfnamefont {J.~A.~M.}\ \bibnamefont
  {Kuipers}},\ and\ \bibinfo {author} {\bibfnamefont {J.~T.}\ \bibnamefont
  {Padding}},\ }\bibfield  {title} {\bibinfo {title} {Non-spherical particles
  in a pseudo-{2D} fluidised bed: {Modelling} study},\ }\href
  {https://doi.org/10.1016/j.ces.2018.08.041} {\bibfield  {journal} {\bibinfo
  {journal} {Chem. Eng. Sci.}\ }\textbf {\bibinfo {volume} {192}},\ \bibinfo
  {pages} {1105} (\bibinfo {year} {2018})}\BibitemShut {NoStop}%
\bibitem [{\citenamefont {Balberg}\ and\ \citenamefont
  {Binenbaum}(1983)}]{Balberg1983}%
  \BibitemOpen
  \bibfield  {author} {\bibinfo {author} {\bibfnamefont {I.}~\bibnamefont
  {Balberg}}\ and\ \bibinfo {author} {\bibfnamefont {N.}~\bibnamefont
  {Binenbaum}},\ }\bibfield  {title} {\bibinfo {title} {Computer study of the
  percolation threshold in a two-dimensional anisotropic system of conducting
  sticks},\ }\href {https://doi.org/10.1103/PhysRevB.28.3799} {\bibfield
  {journal} {\bibinfo  {journal} {Phys. Rev. B}\ }\textbf {\bibinfo {volume}
  {28}},\ \bibinfo {pages} {3799} (\bibinfo {year} {1983})}\BibitemShut
  {NoStop}%
\bibitem [{\citenamefont {Lebovka}\ \emph {et~al.}(2011)\citenamefont
  {Lebovka}, \citenamefont {Karmazina}, \citenamefont {Tarasevich},\ and\
  \citenamefont {Laptev}}]{Lebovka2011}%
  \BibitemOpen
  \bibfield  {author} {\bibinfo {author} {\bibfnamefont {N.~I.}\ \bibnamefont
  {Lebovka}}, \bibinfo {author} {\bibfnamefont {N.~N.}\ \bibnamefont
  {Karmazina}}, \bibinfo {author} {\bibfnamefont {Y.~Y.}\ \bibnamefont
  {Tarasevich}},\ and\ \bibinfo {author} {\bibfnamefont {V.~V.}\ \bibnamefont
  {Laptev}},\ }\bibfield  {title} {\bibinfo {title} {Random sequential
  adsorption of partially oriented linear $k$-mers on a square lattice},\
  }\href {https://doi.org/10.1103/PhysRevE.84.061603} {\bibfield  {journal}
  {\bibinfo  {journal} {Phys. Rev. E}\ }\textbf {\bibinfo {volume} {84}},\
  \bibinfo {pages} {061603} (\bibinfo {year} {2011})}\BibitemShut {NoStop}%
\bibitem [{\citenamefont {Hoshen}\ and\ \citenamefont
  {Kopelman}(1976)}]{Hoshen1976}%
  \BibitemOpen
  \bibfield  {author} {\bibinfo {author} {\bibfnamefont {J.}~\bibnamefont
  {Hoshen}}\ and\ \bibinfo {author} {\bibfnamefont {R.}~\bibnamefont
  {Kopelman}},\ }\bibfield  {title} {\bibinfo {title} {Percolation and cluster
  distribution. {I.} {Cluster} multiple labeling technique and critical
  concentration algorithm},\ }\href {https://doi.org/10.1103/PhysRevB.14.3438}
  {\bibfield  {journal} {\bibinfo  {journal} {Phys. Rev. B}\ }\textbf {\bibinfo
  {volume} {14}},\ \bibinfo {pages} {3438} (\bibinfo {year}
  {1976})}\BibitemShut {NoStop}%
\bibitem [{\citenamefont {van~der Marck}(1997)}]{Marck1997}%
  \BibitemOpen
  \bibfield  {author} {\bibinfo {author} {\bibfnamefont {S.~C.}\ \bibnamefont
  {van~der Marck}},\ }\bibfield  {title} {\bibinfo {title} {Percolation
  thresholds and universal formulas},\ }\href
  {https://doi.org/10.1103/PhysRevE.55.1514} {\bibfield  {journal} {\bibinfo
  {journal} {Phys. Rev. E}\ }\textbf {\bibinfo {volume} {55}},\ \bibinfo
  {pages} {1514} (\bibinfo {year} {1997})}\BibitemShut {NoStop}%
\bibitem [{\citenamefont {Hart}\ and\ \citenamefont
  {Aar{\~a}o~Reis}(2016)}]{Hart2016PRE}%
  \BibitemOpen
  \bibfield  {author} {\bibinfo {author} {\bibfnamefont {R.~C.}\ \bibnamefont
  {Hart}}\ and\ \bibinfo {author} {\bibfnamefont {F.~D.~A.}\ \bibnamefont
  {Aar{\~a}o~Reis}},\ }\bibfield  {title} {\bibinfo {title} {Random sequential
  adsorption of polydisperse mixtures on lattices},\ }\href
  {https://doi.org/10.1103/PhysRevE.94.022802} {\bibfield  {journal} {\bibinfo
  {journal} {Phys. Rev. E}\ }\textbf {\bibinfo {volume} {94}},\ \bibinfo
  {pages} {022802} (\bibinfo {year} {2016})}\BibitemShut {NoStop}%
\bibitem [{\citenamefont {R\'{e}nyi}(1963)}]{Renyi1963}%
  \BibitemOpen
  \bibfield  {author} {\bibinfo {author} {\bibfnamefont {A.}~\bibnamefont
  {R\'{e}nyi}},\ }\bibfield  {title} {\bibinfo {title} {On a one-dimensional
  problem concerning random space filling},\ }\href@noop {} {\bibfield
  {journal} {\bibinfo  {journal} {Selected Translations in Mathematical
  Statistics and Probability}\ }\textbf {\bibinfo {volume} {4}},\ \bibinfo
  {pages} {203} (\bibinfo {year} {1963})}\BibitemShut {NoStop}%
\bibitem [{\citenamefont {Pal{\'a}sti}(1960)}]{Palasti1960}%
  \BibitemOpen
  \bibfield  {author} {\bibinfo {author} {\bibfnamefont {I.}~\bibnamefont
  {Pal{\'a}sti}},\ }\bibfield  {title} {\bibinfo {title} {On some random space
  filling problems},\ }\href@noop {} {\bibfield  {journal} {\bibinfo  {journal}
  {Publications of the Mathematical Institute of the Hungarian Academy of
  Sciences}\ }\textbf {\bibinfo {volume} {5}},\ \bibinfo {pages} {353}
  (\bibinfo {year} {1960})}\BibitemShut {NoStop}%
\bibitem [{\citenamefont {Az\'ema}\ and\ \citenamefont
  {Radja\"{\i}}(2010)}]{Azema2010}%
  \BibitemOpen
  \bibfield  {author} {\bibinfo {author} {\bibfnamefont {E.}~\bibnamefont
  {Az\'ema}}\ and\ \bibinfo {author} {\bibfnamefont {F.}~\bibnamefont
  {Radja\"{\i}}},\ }\bibfield  {title} {\bibinfo {title} {Stress-strain
  behavior and geometrical properties of packings of elongated particles},\
  }\href {https://doi.org/10.1103/PhysRevE.81.051304} {\bibfield  {journal}
  {\bibinfo  {journal} {Phys. Rev. E}\ }\textbf {\bibinfo {volume} {81}},\
  \bibinfo {pages} {051304} (\bibinfo {year} {2010})}\BibitemShut {NoStop}%
\end{thebibliography}%

\end{document}